\documentclass[sigconf, nonacm]{acmart}
\usepackage{subcaption}
\usepackage{tikz}
\usepackage{array,multirow,graphicx}
\usepackage{listings}

\lstdefinelanguage{SQLPP}{
  alsoletter={-"'_:\$[]()0123456789\\},
  morekeywords={
    SELECT,
    FROM,
    WHERE,
    GROUP,
    BY,
    AS,
    by,
    with,
    order,
    desc,
    asc,
    limit,
    replace,
    delete,
    insert,
    into,
    some,
    every,
    satisfies,
    and,
    or,
    use,
    dataverse,
    set,
    simfunction,
    simthreshold,
    create,
    index,
    type,
    fuzzy,
    keyword,
    NGram,
    rtree,
    VALUE,
    WITH,
    ORDER,
    LIMIT,
    SOME,
    EVERY,
    SATISFIES,
    IN,
    DESC,
    ASC,
    AND,
    UNNEST,
    LET,
    DISTINCT
  },
  basicstyle=\sf,
  boxpos=centered,
  escapeinside={\%*}{*)},
  keywordstyle={\sf\textbf},
  identifierstyle=\texttt,
  commentstyle=\textit,
  literate={<=}{{\litleq}}1 {>=}{{\litgeq}}1,
  literate={~=}{$\sim{}=$}1 
}[keywords,comments,strings]

\lstdefinelanguage{AQLSchema}{
  alsoletter={-"'_:(\$\\},
  morekeywords={
    use,
    declare,
    type,
    as,
    open,
    closed,
    dataset,
    dataverse,
    nodegroup,
    on,
    partitioned,
    by,
    primary,
    key,
    create,
    index,
    bigint,
    double,
    date,
    polygon,
    with,
    filter,
    CREATE,
    TYPE,
    string,
    int,
    AS,
    DATASET,
    PRIMARY,
    KEY,
    CLOSED,
    OPEN,
    WITH,
    true
  },
  basicstyle=\sf,
  keywordstyle={{\sf\textbf}},
  identifierstyle=\texttt,
  commentstyle=\textit,
  literate={<=}{{\litleq}}1 {>=}{{\litgeq}}1
}[keywords,comments,strings]

\lstdefinelanguage{AQLSchema2}{
  alsoletter={-"'_:(\$\\},
  morekeywords={
    use,
    declare,
    type,
    as,
    open,
    closed,
    dataset,
    dataverse,
    nodegroup,
    on,
    partitioned,
    by,
    primary,
    key,
    create,
    index,
    int32,
    int64,
    string,
    datetime,
    double,
    date,
    point,
    with,
    filter,
    TwitterUserType
  },
  basicstyle=\sf,
  keywordstyle=\textbf,
  identifierstyle=\texttt,
  commentstyle=\textit,
  literate={<=}{{\litleq}}1 {>=}{{\litgeq}}1
}[keywords,comments,strings]

\lstnewenvironment{aql}
   {\lstset{language=AQL,basicstyle=\small,basewidth={.5em,.5em}}}
   {}
   
\lstnewenvironment{sqlpp}
   {\lstset{language=SQLPP,basicstyle=\hspace{4cm}\small,basewidth={.6em,.7em}}}
   {}

\lstnewenvironment{aqlscriptsize}
   {\lstset{language=AQL,basicstyle=\scriptsize,basewidth={.5em,.5em},numbers=left}}
   {}

\lstnewenvironment{aqltiny}
   {\lstset{language=AQL,basicstyle=\tiny,basewidth={.5em,.5em},numbers=left}}
   {}

\lstnewenvironment{aqlschema}
   {\lstset{language=AQLSchema,basicstyle=\small,basewidth={.5em,.5em}}}
   {}
   
\lstnewenvironment{inlineaqlschema}
   {\lstset{language=AQLSchema,basicstyle=\noindent\small,basewidth={.5em,.5em}}}
   {}
   
\lstnewenvironment{inlineaqlschema2}
   {\lstinline{language=AQLSchema,basicstyle=\noindent\small,basewidth={.5em,.5em}}\lstMakeShortInline[columns=fixed]|}
   {}
   
\lstnewenvironment{aqlschema2}
   {\lstset{language=AQLSchema2,basicstyle=\small,basewidth={.5em,.5em}}}
   {}
   
\lstnewenvironment{scala}
   {\lstset{language=SCALA, showstringspaces=false,tabsize=3,basicstyle=\linespread{1}\small\ttfamily,basewidth={.5em,.5em}}}
   {}

\newcommand\vldbdoi{XX.XX/XXX.XX}

\newcommand\vldbavailabilityurl{}

\newcommand{\pgftextcircled}[1]{
    \setbox0=\hbox{#1}%
    \dimen0\wd0%
    \divide\dimen0 by 2%
    \begin{tikzpicture}[baseline=(a.base)]%
        \useasboundingbox (-\the\dimen0,0pt) rectangle (\the\dimen0,1pt);
        \node[circle,draw,outer sep=0pt,inner sep=0.1ex, fill=black] (a) {\textcolor{white}{#1}};
    \end{tikzpicture}
}
\let\textcircled=\pgftextcircled

\newcommand{\pgfbluecircle}[1]{
    \setbox0=\hbox{#1}%
    \dimen0\wd0%
    \divide\dimen0 by 2%
    \begin{tikzpicture}[baseline=(a.base)]%
        \useasboundingbox (-\the\dimen0,0pt) rectangle (\the\dimen0,1pt);
        \node[circle,draw,outer sep=0pt,inner sep=0.1ex, fill=blue] (a) {\textcolor{white}{#1}};
    \end{tikzpicture}
}
\let\bluecircle=\pgfbluecircle

\newcommand{\pgfredcircle}[1]{
    \setbox0=\hbox{#1}%
    \dimen0\wd0%
    \divide\dimen0 by 2%
    \begin{tikzpicture}[baseline=(a.base)]%
        \useasboundingbox (-\the\dimen0,0pt) rectangle (\the\dimen0,1pt);
        \node[circle,draw,outer sep=0pt,inner sep=0.1ex, fill=red] (a) {\textcolor{white}{#1}};
    \end{tikzpicture}
}
\let\redcircle=\pgfredcircle

\begin{document}
\title{Columnar Formats for Schemaless LSM-based Document Stores}

\author{Wail Y. Alkowaileet}
\affiliation{%
  \institution{University of California, Irvine}
  \city{Irvine}
  \state{CA}
}
\email{w.alkowaileet@uci.edu}

\author{Michael J. Carey}
\affiliation{%
  \institution{University of California, Irvine}
  \city{Irvine}
  \state{CA}
}
\email{mjcarey@ics.uci.edu}

\begin{abstract}
\label{abstract}

In the last decade, document store database systems have gained more traction for storing and querying large volumes of semi-structured data. However, the flexibility of the document stores' data models has limited their ability to store data in a columnar-major layout --- making them less performant for analytical workloads than column store relational databases. In this paper, we propose several techniques, based on piggy-backing on Log-Structured Merge (LSM) tree events and tailored to document stores to store document data in a columnar layout. We first extend the Dremel format, a popular on-disk columnar format for semi-structured data, to comply with document stores' flexible data model. We then introduce two columnar layouts for organizing and storing data in LSM-based storage. We also highlight the potential of using query compilation techniques for document stores, where values' types are known only at runtime. We have implemented and evaluated our techniques to measure their impact on storage, data ingestion, and query performance in Apache AsterixDB. Our experiments show significant performance gains, improving the query execution time by orders of magnitude while minimally impacting ingestion performance.
\end{abstract}

\maketitle


\ifdefempty{\vldbavailabilityurl}{}{
\vspace{.3cm}
\begingroup\small\noindent\raggedright\textbf{PVLDB Artifact Availability:}\\
The source code, data, and/or other artifacts have been made available at \url{\vldbavailabilityurl}.
\endgroup
}

\section{Introduction}
\label{intro}

In recent years, columnar storage systems have been widely adopted in data warehouses for analytical workloads, where typical queries access only a few fields of each tuple. By storing columns contiguously as opposed to rows, column store systems only need to read the columns involved in a query and the IO cost becomes significantly smaller compared to reading whole tuples~\cite{c-store, dremel}. As a result, open source and commercial relational column-store systems such as MonetDB \cite{monetdb, x100} (and the commercial version Actian Vector \cite{vector}), and C-Store \cite{c-store} (commercialized as Vertica \cite{vertica}) have gained more popularity as data warehouse solutions.

For semi-structured data, Dremel \cite{dremel} and its open source implementation Apache Parquet \cite{parquet} offer a way to store homogenous JSON-like data in a columnar format. Apache Parquet has become the de facto file format for popular big data systems such as Apache Spark and even for ``smaller'' data processing libraries like Python's Panda. However, storing data in a column-oriented fashion for document store systems such as MongoDB \cite{mongodb}, Couchbase Server ~\cite{couchbase} or, Apache AsterixDB \cite{asterixdb, asterix-overview-vldb2015, asterixdb-midflight} is more challenging, as: \textbf{(1)} Declaring a schema before loading or ingesting data is not required in document store systems. Thus, the number of columns and their types are determined upon data arrival. \textbf{(2)} Document store systems do not prohibit a field from having two or more different types, which adds another layer of complexity. Despite the performance gains of columnar formats, many users with big data still choose the flexibility offered by document stores. 

Many prominent document stores, such as MongoDB and Couchbase Server, adopt Log-Structured Merge (LSM) trees \cite{lsm} in their storage engines for their superior write performance. LSM lifecycle events (mainly the flush operations) allow transforming the ingested records upon writing them to disk. This paper proposes several techniques to overcome document challenges to allow storing and querying semi-structured data in a columnar format for LSM-based document stores. We first extend the Dremel format to comply with document stores' flexible data model, which permits values with heterogeneous types. We then use the same techniques proposed in the tuple compactor framework \cite{compact} to exploit the LSM flush operation to infer the schema and write the records (initially in row format) as columns using the extended Dremel format. 


We present two new models in our work here for storing columns in an LSM B$^+$-tree index. In the first model, we store columns using a Partitioned Attributes Across (PAX)-like \cite{pax} format, where each column occupies a contiguous region (called a minipage) within a B$^+$-tree's leaf page. We refer to this model as the AsterixDB PAX model or APAX for short. In the second model, we stretch the PAX minipages to become megapages, where a column could occupy multiple pages. We refer to this model as the AsterixDB Mega-Attributes Across (AMAX). Despite their names, these layouts are not system-specific and should only require a few modifications to be adopted by other LSM-based document stores. We evaluate and show the pros and cons of the two approaches in terms of (1) ingestion performance, (2) query performance, and (3) memory and CPU consumption.

The goal of continuously reducing the I/O cost in disk-based databases is objectively justified (and is a focus in this paper). However, with the ever-growing advancements in storage technologies, the role of CPU cost becomes even more apparent. In our evaluation, we observe an interesting phenomenon in certain types of workloads, where we have been able to reduce the size of the data needed to process a query by several factors using the APAX and AMAX formats as compared to the vector-based format (a row-major format) from \cite{compact}. However, the associated improvement to query execution time in certain cases was negligible due to increased CPU cost. The dominant factor determining the CPU cost is the query execution model. Modern Database Management Systems (DBMSs) have moved away from using the traditional iterator model \cite{ibm-iter, volcano-iter} to use other execution models (such as the batch model \cite{block} and the materialization model \cite{materialzation}) to minimize the CPU overhead. However, hand-written code outperforms all three models \cite{x100}. Thus, code generation and query compilation have become a major contributors to the performance gains of many recent data processing engines \cite{spark-compiler, drill} and DBMSs \cite{peloton-qe, neumann} alike.

In this work, We shed light on the possibility of using query compilation techniques for document stores, where value types are not known until runtime. We utilize the Oracle Truffle framework (called Truffle hereafter) \cite{truffle} to implement an internal language for processing data stored in a Java-based document store. 
Even though we only translate part of a query plan, our evaluations show a tremendous improvement over AsterixDB's existing model.

To show their benefits, we have implemented the proposed techniques to store document data in a columnar format and produce a compiled query plan in Apache AsterixDB. This enabled us to conduct an extensive evaluation of the APAX and AMAX formats and present their tradeoffs for different datasets. We also show the impact of utilizing Truffle to generate and execute queries against different datasets stored as AMAX and APAX, as well as the original schemaless row format of AsterixDB and the recently proposed Vector-based format.
%

\section{Background}
\label{background}

\subsection{Apache AsterixDB}
AsterixDB is a parallel semi-structured Big Data Management System (BDMS) that runs on large, shared-nothing, commodity computing clusters. To prepare the reader, here we give a brief overview of AsterixDB's storage engine \cite{storage}, its compiler, Algebricks \cite{algebricks}, and its query execution engine, Hyracks \cite{hyracks}.

\subsubsection{\textbf{Storage Engine:}}
\label{sec:asterix}
\begin{figure}[h]
  \includegraphics[width=0.43\textwidth]{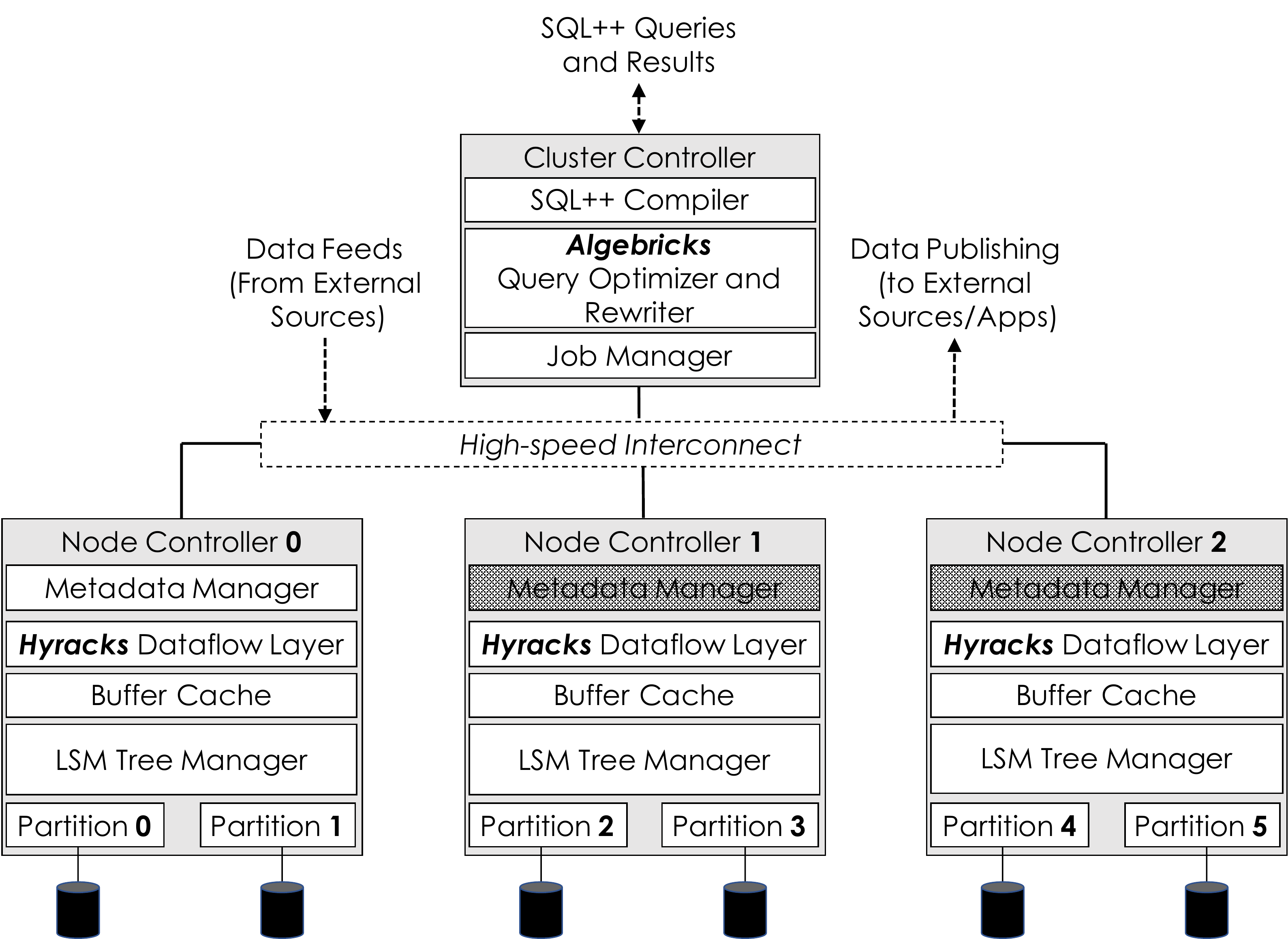}
  \caption{Apache AsterixDB architecture}
  \label{fig:asterix-arch}
\end{figure}

An AsterixDB cluster consists of worker nodes called Node Controllers (NCs) managed by a Cluster Controller (CC) node. Figure \ref{fig:asterix-arch} shows an AsterixDB cluster with three NCs, each with two data partitions that store data on two separate storage devices. Data partitions within the same NC (e.g., Partition 0 and Partition 1 in NC0) share the same resources (e.g., memory budget) configured for each NC; however, the data stored in each partition is managed independently.

\begin{figure}[h]
\centering
  \begin{subfigure}[b]{0.49\textwidth}
  	  \includegraphics[width=\textwidth]{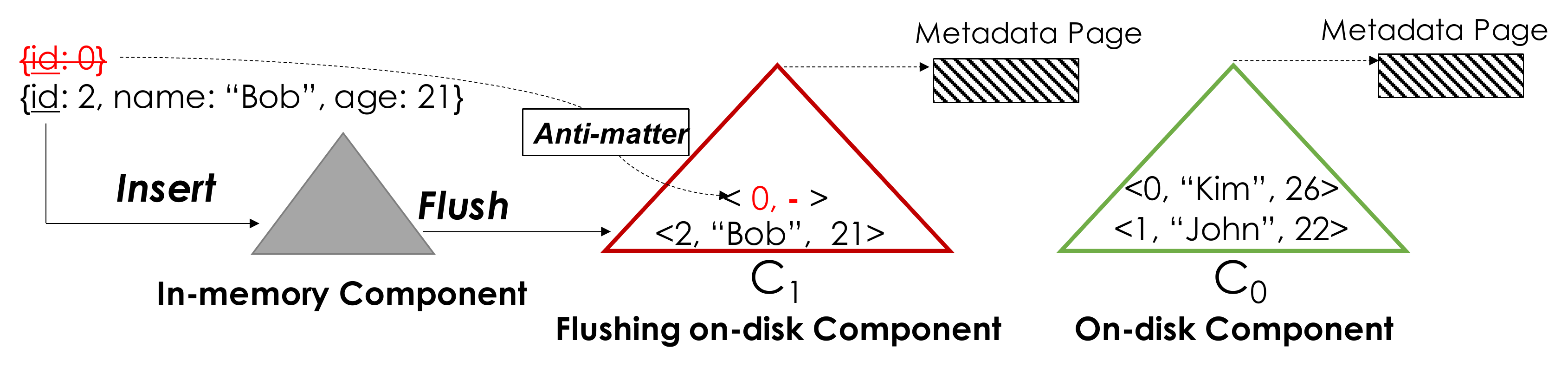}
  	  \caption{}
  	  \label{fig:flush}
  \end{subfigure}
    \begin{subfigure}[b]{0.49\textwidth}
  	  \includegraphics[width=\textwidth]{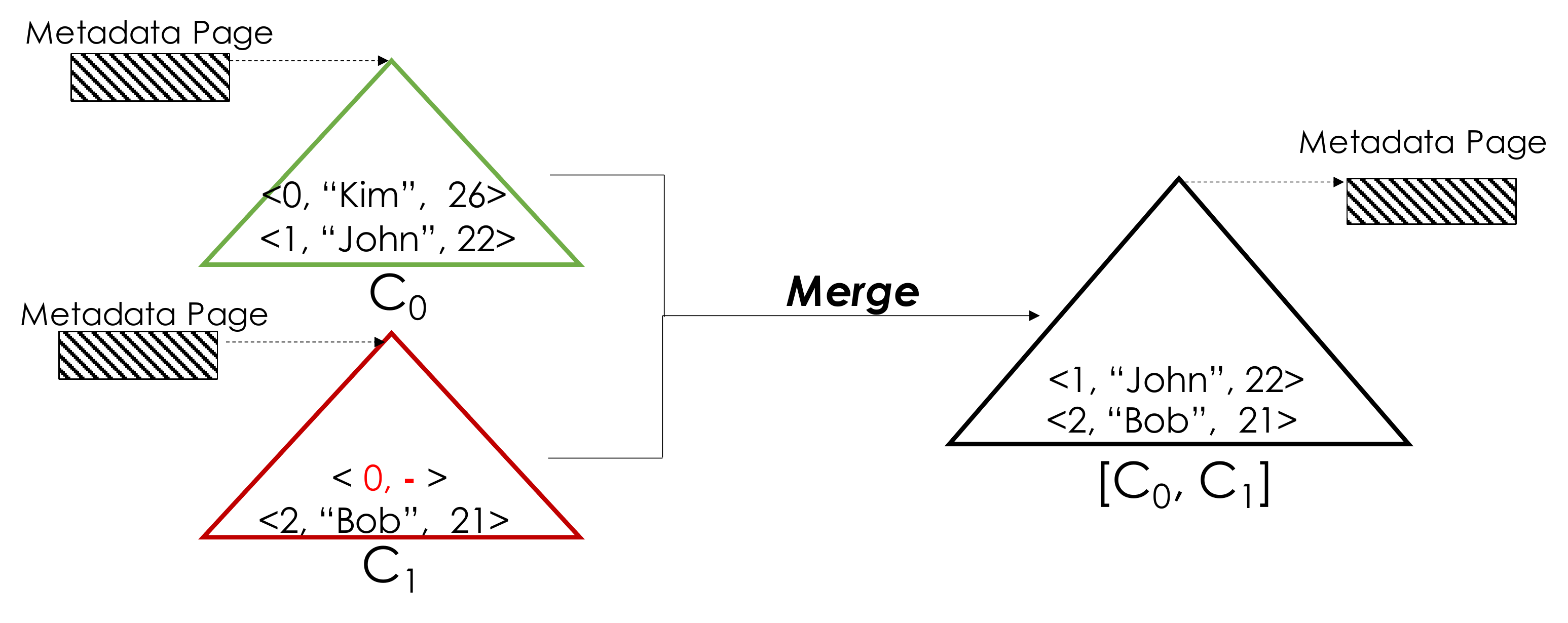}
      \vspace{-0.6cm}
  	  \caption{}
  	  \label{fig:merge}
  \end{subfigure}
  \caption{(a) Flushing component $C_1$ (b) Merging the two components $C_0$ and $C_1$ into a new component $[C_0, C_1]$}
  \vspace{-0.3cm}
  \label{fig:lsm-lifecycle}
\end{figure}

AsterixDB stores its datasets' records, spread across the data partitions in all NCs, in primary LSM B$^+$-tree indexes. Newly inserted records are hash-partitioned using their primary key(s) into the data partitions (Partition 0 to Partition 5 in Figure \ref{fig:asterix-arch}) and inserted into the resulting partition's primary LSM in-memory component. When the in-memory component is full, the LSM Tree Manager flushes the in-memory component's records into a new LSM on-disk component on the partition's storage device, as shown in Figure \ref{fig:flush}. Upon completion, the tree manager marks the flushed component as valid by setting a validity bit on the component's metadata page and freeing the in-memory component to serve subsequent inserts. LSM on-disk components are immutable and, hence, updates and deletes are both handled by inserting new entries. A delete operation adds an ``anti-matter'' (or tombstone) entry to indicate that a record with a specified key has been deleted. An update simply adds a new record with the same key as the original one. As on-disk components accumulate, the tree manager periodically merges them into larger components in the background according to a configured merge policy \cite{storage,luo2019efficient}, which determines when and what to merge. Deleted and old updated records are garbage-collected during the merge operation. In Figure \ref{fig:merge}, during the merge of C0 and C1 (from Figure \ref{fig:flush}) into a new disk-component (called [C0, C1]), the record with id = 0  and its corresponding anti-matter annihilate each. On completion, the older on-disk components (C0 and C1) are deleted and replaced by the newly created component [C0, C1].

\subsubsection{\textbf{Query Execution Model:}}
To query the data stored in AsterixDB, a user can submit a query written in SQL++ \cite{sql++,sql++_book} to the CC, which generates an optimized query plan and then compiles it into a Hyracks job. The compiled Hyracks job is then distributed to the query executors in all partitions to run in parallel. Hyracks jobs consist of operators and connectors \cite{hyracks}, where data flows between operators over connectors as a batch of tuples. Each batch of tuples received by an operator is processed and passed over to the next operator as a new batch. 

\subsection{LSM-based Tuple Compaction Framework}
\label{sec:compactor}
The flexibility of document stores is targeted for applications where the schema can change without human intervention. However, document stores' flexibility is not free, as each record stores its schema instead of storing it in a centralized catalog. In a previous work \cite{compact}, we presented a Tuple Compactor framework (implemented in Apache AsterixDB) that addresses this issue by exploiting LSM lifecycle events to infer the component's schema and compact its records using the inferred schema. 


\begin{figure}[h]
  \includegraphics[width=0.48\textwidth]{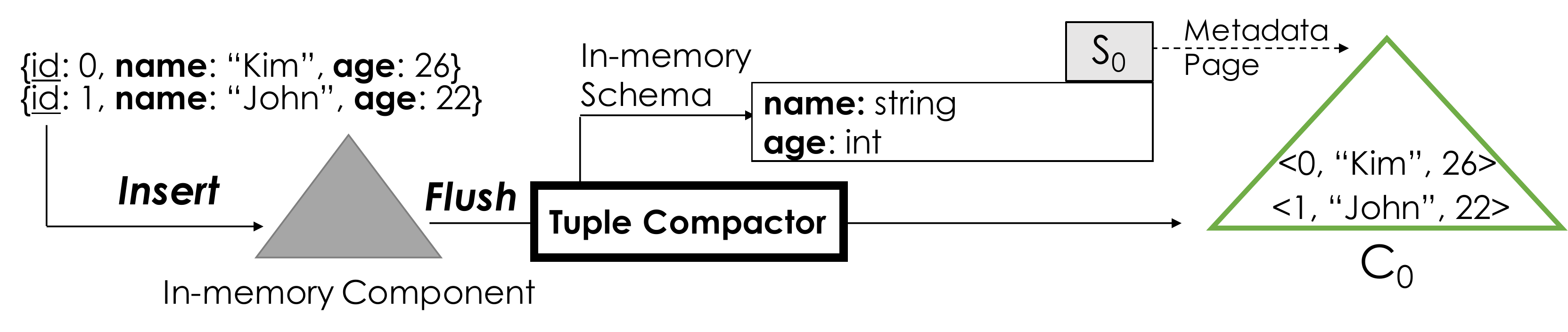}
  \vspace{-0.5cm}
  \caption{Schema inference workflow}
  \label{fig:compaction}
\end{figure}

To illustrate, when creating a dataset in AsterixDB, each partition in every NC (Figure \ref{fig:asterix-arch}) starts with an empty dataset. During data ingestion, each partition inserts the received records into the in-memory component as in normal operation. When the memory component is full, the in-memory components' records are flushed into a new on-disk component, during which time the tuple compactor takes this opportunity to infer the schema and compact the flushed records. Figure \ref{fig:compaction} depicts the workflow of the tuple compactor along with the inferred schema. In the figure, we see that the tuple compactor has inferred two fields, name and age, with the types string and integer, respectively, from the flushed records. Upon completing the flush operation, the inferred schema is persisted into the component's metadata page. Subsequent flushes follow the same workflow to build the schema for all of the ingested records. Among the flushed components, the schema of the latest flush is always a super-set of all previous schemas. Thus, we only persist the most recent component's schema into a merged component's metadata page during a merge operation.

Also in \cite{compact}, we introduced the Vector-based format --- a \sloppy{non-recursive}, compaction-friendly, physical data format for storing semi-structured data. The vector-based format's main idea is to separate the data values from the records' metadata, which describes the record's structure. This separation enables the tuple compactor to efficiently operate on the record's metadata during the schema inference and record compaction processes. Additionally, being a non-recursive, the vector-based format allows the tuple compactor to iteratively process a record, which is more cache-friendly than AsterixDB's recursive format \cite{asterixdb-serialization}. We refer interested readers to \cite{compact, extended} for more details about the vector-based format.

\begin{figure*}[t]
    \begin{subfigure}[b]{0.42\textwidth}
        \includegraphics[width=\textwidth, height=0.56\textwidth]{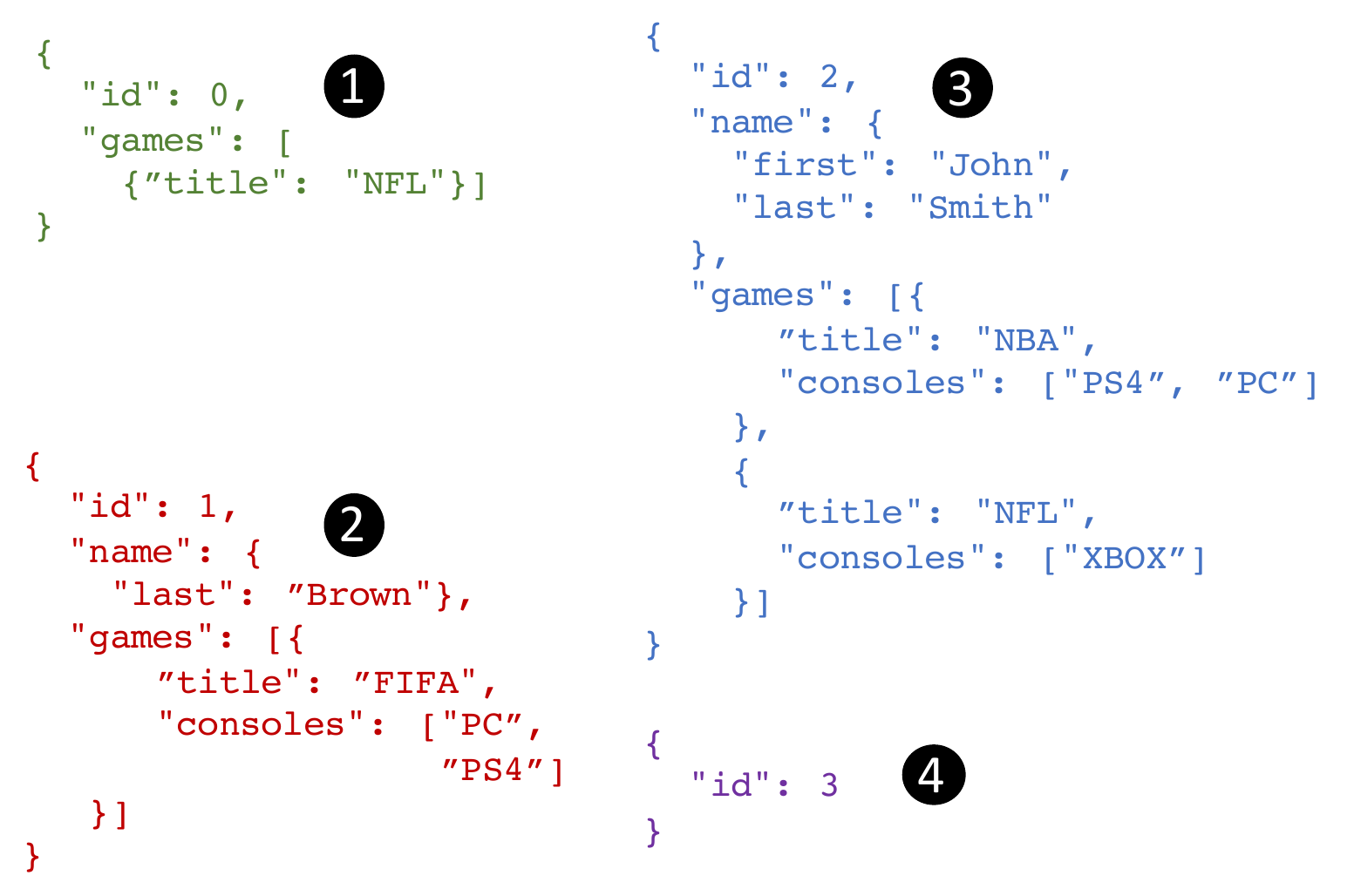}
        \caption{}
        \label{fig:dremel-records}
    \end{subfigure}
    \begin{subfigure}[b]{0.56\textwidth}
        \includegraphics[width=\textwidth, height=0.5\textwidth]{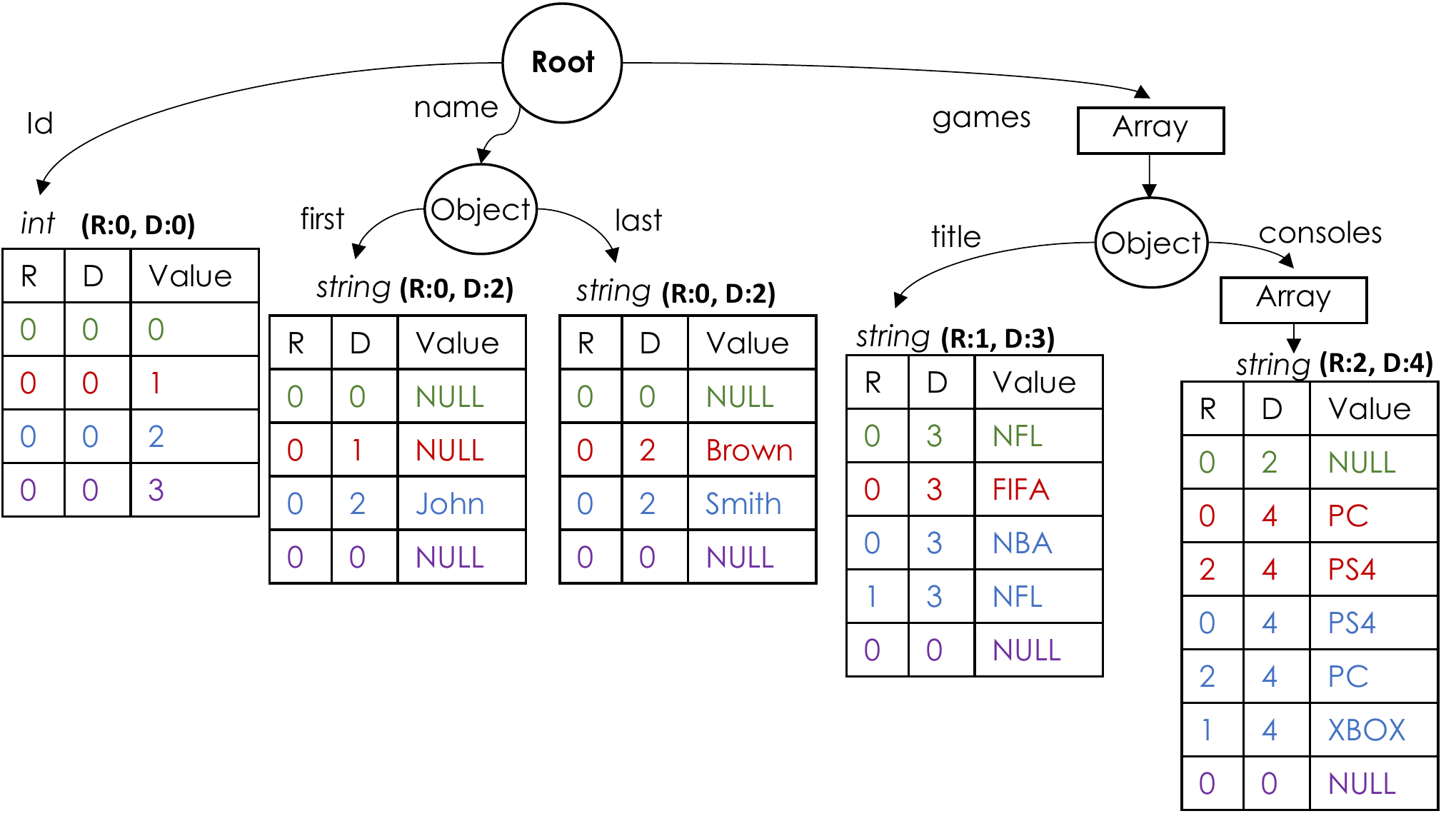}
        \caption{}
        \label{fig:dremel-columns}
    \end{subfigure}
    \caption{(a) Raw JSON records (b) Dremel declared schema along with the column representation of the raw JSON records}
    \label{fig:dremel}
\end{figure*}
\section{A Flexible Columnar Format for nested semi-structured data}
\label{dremel}
Inferring the schema and compacting schemaless semi-structured records, using the tuple compactor framework \cite{compact}, reduces their overall storage overhead and consequently improves the query execution time. However, these compacted records are still in a row-major format, which is less than ideal for analytical workloads as compared to columnar formats. One of the main reasons that document stores do not support storing data in a columnar format is the flexibility of their data model. E.g., the Dremel (or Parquet) format still requires a schema that describes all fields to be declared a priori, and all field values must conform to one type. In this section, we give a brief overview of the Dremel format. Then, we present our extensions to Dremel to allow schema changes, such as adding new values and changing their types.
\vspace{-0.3cm}

\subsection{Dremel Format}
The Dremel format allows for storing nested records in a columnar fashion, where atomic values of different records are stored contiguously in chunks. For better illustration, Figure \ref{fig:dremel} shows an example of four JSON records (Figure \ref{fig:dremel-records}) about video gamers stored in Dremel format along with the schema (Figure \ref{fig:dremel-columns}). The schema's inner nodes represent the nested values (objects and arrays), whereas the leaf nodes represent the atomic values such as integers and strings. The schema describes the JSON records structure, where the root has three fields $id$, $name$, and $games$ with the types integer, object, and array, respectively. The $name$ object consists of $first$ and $last$ name pairs, both of which are of type string. Next is the array of objects $games$, which stores information about the gamers' owned games, namely the games' $titles$ and the different versions of a game the gamers own for different $consoles$. Every value (nested or atomic) in our example is optional except for the record's key $id$. The optionality of all non-key values is synonymous with the schemaless document store model, which is the scope of this paper. We encourage interested readers to refer to \cite{dremel, dremel-made-simple} for more details on the representation of non-optional values.
 
 The tables underneath the schema's leaf nodes shown in Figure \ref{fig:dremel-columns} depict the Dremel's columnar-striped representation of the records' atomic values from Figure \ref{fig:dremel-records}. Each table consists of three columns: \textbf{R}, \textbf{D}, and \textbf{Value}, where \textbf{R} and \textbf{D} denote the Repetition-Level and Definition-Level of each Value as presented in \cite{dremel}. The definition levels' roles determine the NULL occurrence level for a nested value. The repetition levels' roles determine the start and end of a repeated value (array). The pairs (R:$x$, D:$y$), shown at the top of each table in Figure \ref{fig:dremel-columns}, indicate the maximum value for the repetition and definition levels for each atomic value.
 
To explain, the column $name.first$ has a maximum repetition level 0 indicating a non-repeated value (or not an array element), whereas the definition level 2 is the level of the leaf node in the schema's tree ($root$ (0) $\rightarrow$ $name$ (1) $\rightarrow$ $first$ (2)). In the first record \textcircled{1} in Figure \ref{fig:dremel-records}, the definition level for the value $name.first$ is 0, which indicates that only the root was present in the path $root \rightarrow name \rightarrow first$. Thus, the value for the $name.first$ in the first record in our example is NULL. For the second record \textcircled{2}, the definition level for the $name.first$ value is 1, which indicates the the $name$ object is present but not the atomic value $first$. In record \textcircled{3}, the gamer's $first$ name is "John"; hence, it has definition level 2. In the last record \textcircled{4}, the $name$ value is missing, indicated by the definition level 0. The key field $id$ is required for all records, and the key's definition level value is ignored. Thus, the $id$'s maximum definition level is 0.
 
 For repeated values (array elements) such as the games' $titles$ column (denoted as $games[*].title$) in our example, the repetition levels determine the array starts and ends for each record. Note that for the repeated values $games[*].title$, the maximum repetition and definition levels are 1 and 3, respectively. The first record has only one value (0, 3, "NFL"), where the triplet (r, d, v) denotes its repetition level (r), definition level (d), and value (v), respectively. The repetition level 0 indicates that the value "NFL" is the record's first $games[*].title$ repeated value, and the definition level 3 indicates that the value is present. The following value (0, 3, "FIFA) corresponds to the second record, as the repetition level 0 indicates that the current value is, again, the first $games[*].title$ repeated value. Similarly, the value (0, 3, "NBA") corresponds to the third record; however, the following value (1, 3, "NFL") also corresponds to the same record (i.e., the second element of the array), which is indicated by repetition level 1. Whenever a value's repetition level is equal to the column's maximum repetition level, we know that it is the $i^{th}$ repeated value (or $i^{th}$ element of the array), where $i > 0$. In the last record, the array $games$ itself is missing, and thus its definition level is 0.
 
 In Figure \ref{fig:dremel-columns}, the array $consoles$ (whose full path is denoted as $games[*].consoles[*]$) is an ancestor of the outer array $games$. In other words, the atomic values of $consoles$ belong to \textbf{two} nested arrays $games$ and $consoles$. Therefore, the maximum repetition level for the column $games[*].consoles[*]$ is 2. Like in the column $games[*].title$, the value (0, 2, NULL) corresponds to the first record, as indicated by the repetition level 0. However, the first record is missing the field consoles; thus, its definition level 2 indicates that the consoles array is missing, but its parent is present. The following value (0, 4, "PC") is the first $games[*].consoles[*]$'s value for the second record, as indicated by its repetition level 0. The definition level 4 here means that the value is present and the value is "PC". The next value (2, 4, "PS4") has a repetition level 2, the maximum repetition level for the column $games[*].consoles[*]$, which means it is the second value of the array consoles. Likewise, the following two values (0, 4, "PS4") and (2, 4, "PC") correspond to the third record, and both constitute the first and second elements of the first consoles array. The following value's (1, 4, "XBOX") repetition level 1 means it still corresponds to the same record; however, the value marks the beginning of the record's second $consoles$' array, which has a single element "XBOX". The last value (0, 0, NULL) indicates that the array $games$ is missing from the last record.
 
 \subsection{Extended Dremel Format}
 \label{sec:extended-dremel}
The Dremel format is designed for storing semi-structured data such as JSON documents in a columnar-oriented fashion. However, the requirement of declaring the schema a priori prohibits flexible document stores from adopting it. Additionally, the Dremel format does not support union types (at the time of writing this paper), where a value (or a column) can be of different types in different records. In this work, we present our extensions to the Dremel format to accommodate document stores flexible data model.

\subsubsection{\textbf{Array Values:}}
Earlier, we explained the Dremel approach for representing repeated values (arrays) using repetition levels. Here we observe that the repetition levels (i) are redundant among nodes that share the same ancestor array, and (ii) along with the definition levels, they occupy more bits than needed to represent optional repeated values. 

In our example, for (i), notice how the repetition levels of the column $games[*].name$ is a subset of the column $games[*].consoles[*]$'s repetition levels (redundancy), as both share the same array ancestor $games[*]$. The entire repetition levels of the column $games[*].title$  [0, 0, 0, 1, 0] appear in the same order as the column $games[*].consoles[*]$'s repetition levels [\textbf{0}, \textbf{0}, 2, \textbf{0}, 2, \textbf{1}, \textbf{0}]. For (ii), we observe that all values with repetition levels greater than 0 must have definition levels greater or equal to the array's level. Recall that a value with a repetition level greater than 0 corresponds to the $i^{th}$ element of an array, where $i > 0$. When the repetition level is greater than 1, it implies that an array exists and that its length is greater than one. Also recall that when the definition level is smaller than an array's level in the schema, it means that the array itself is NULL. As a consequence, having a repetition level greater than 0 and a definition level smaller than the array's level would be contradictory. It would mean the array exists and that its length is greater than one, but that the array itself is NULL (or does not exist). Given that, the number of bits for both the definition and repetition levels is more than what is needed to represent repeated values. 

\begin{figure}[ht]
    \centering
    \includegraphics[width=0.39\textwidth]{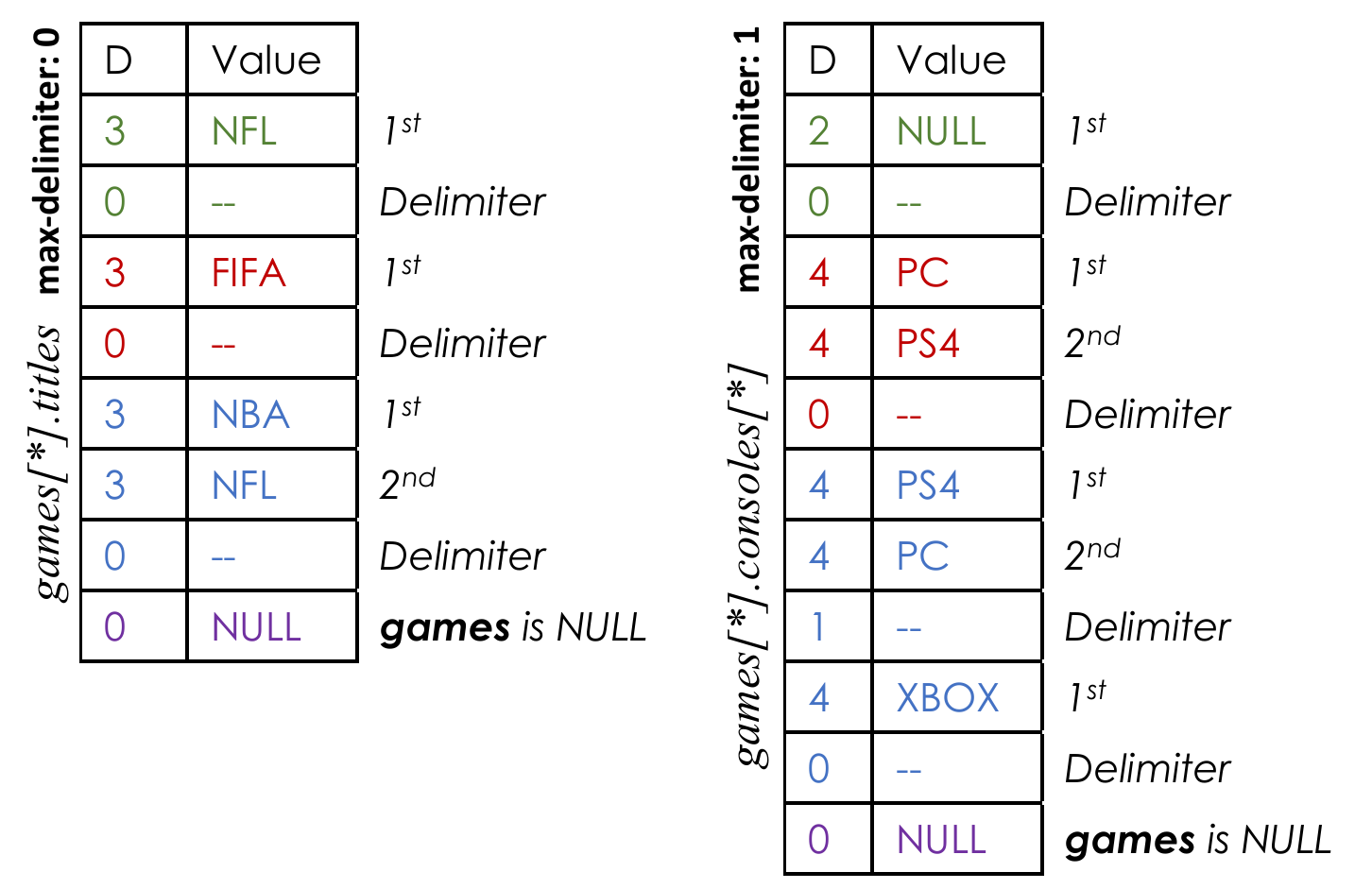}
    \vspace{-0.3cm}
    \caption{Delimited repeated values for records in Figure \ref{fig:dremel-records}}
    \vspace{-0.2cm}
    \label{fig:extended-array}
\end{figure}

For these reasons, we will adopt a different approach for representing repeated values without repetition levels. Recall (ii), which says that the i$^{th}$ definition level (where $i>0$) of a repeated value cannot be smaller than the array's definition level. Thus, we can use such definition level values as delimiters instead of repetition levels. To illustrate, Figure \ref{fig:extended-array} shows the values of both the $games[*].title$ and $games[*].conosoles[*]$ columns for the records in Figure \ref{fig:dremel-columns} using the proposed approach. The repeated values in both columns have the same definition levels as in the original Dremel format. However, the definition levels also indicate end-of-array by having a delimiter value. In the column $games[*].title$, the first three records' repeated values are delimited by the definition level 0, which is the \textit{max-delimiter} value in the column as shown in Figure \ref{fig:dremel-columns}. The value that follows a delimiter indicates the start of the next array, and the value itself is the array's first value --- except for the last repeated value in Figure \ref{fig:extended-array}, where the definition level 0 that indicates the array $games$ is NULL in the last record. Note that the last value's definition level of 0 cannot be a delimiter since it is the first value after the preceding delimiter.

In the case of nested arrays, as in the column $games[*].consoles[*]$, the max-delimiter is 1, which indicates that the two delimiter values 0 and 1 are for the outer ($games$) and inner ($consoles$) arrays, respectively. The first value in the column $games[*].consoles[*]$ is NULL with definition level 2, which indicates that the outer array $games$ is present but its sole element is NULL. The following value is a delimiter of the outer array $games$, indicated by the definition level 0. The next two values are the first and second array elements of the second record's array $consoles$, followed by a delimiter with the definition level 0. We omit the definition level 1, the delimiter for the inner array consoles, since the delimiter 0 also encompasses the inner delimiter 1. The next two values are the first and second $consoles$ array elements in the third record. The next delimiter of 1 here indicates the end of the first consoles array ["PS4", "PC"], and the next value marks the start of the second $consoles$ array ["XBOX"] in the same record. The following delimiter 0 indicates the end of the repeated values in the third record. In the last value, the definition level of 0 implies that the $games$ array is NULL in the last record. Like the last value of the column $games[*].title$, the definition level 0 here is not a delimiter as it is the first value after the last delimiter.

\subsubsection{\textbf{Schema Changes and Heterogeneous Values:}}

For LSM-based document stores, one could use the approach proposed in~\cite{compact} (summarized in Section \ref{sec:compactor}) to obtain the schema and use it to dissect the values into columns. However, a major challenge for supporting columnar formats in document stores is handling their potentially heterogeneous values. For example, the two records \lstinline[language=AQLSchema,basicstyle=\noindent\small]|{"id": 1, "age": 25}| and \lstinline[language=AQLSchema,basicstyle=\noindent\small]|{"id": 2, "age": "old"}| are valid records and both could be stored in a document store. Similarly, document stores allow storing an array that consists of heterogeneous values, such as the array \lstinline[language=AQLSchema,basicstyle=\noindent\small]|[0, "1", {"seq": 2}]|. Limiting the support for storing data in a columnar format to datasets with homogeneous values is maybe enough for most cases, as evidently shown by the popularity of Parquet. However, including support for datasets with heterogeneous values is a desired feature for certain use cases, especially when the users have no control over how the data is structured, like when ingesting data from web APIs \cite{schema-change1, schema-change2}. In this section, we detail our approach for handling schema changes and heterogeneous types.

\begin{figure}[ht]
    \centering
    \includegraphics[width=0.48\textwidth, height=0.16\textwidth]{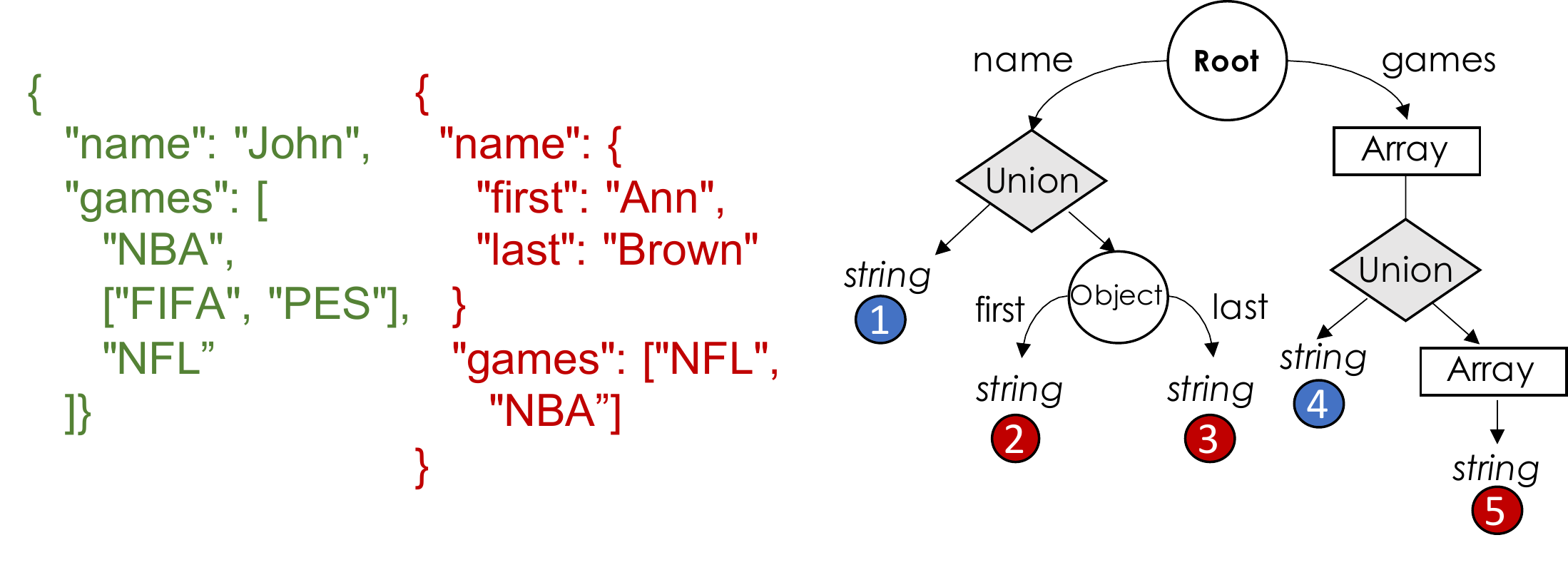}
    \vspace{-0.5cm}
    \caption{Example of heterogeneous values and their schema}
    \label{fig:extended-union1}
    \vspace{-0.5cm}
\end{figure}

In our previous work \cite{compact}, we introduced union types in our inferred schemas to represent values with heterogeneous types. Figure \ref{fig:extended-union1} depicts an example of two variant records with their inferred schema. The inferred schema shows that the records have different types for the same value. The first is the field $name$, which could be a string or an object. Thus, we infer that the $name$'s type is a union of string and object. The second union type corresponds to the $games$ array's elements, where each element could be of type string or array of strings. In the schema, we observe that union nodes resemble a special case of $object$ nodes, where the keys of the union nodes' children are their types. For example, the union node of the field $name$, in the schema shown in Figure \ref{fig:extended-union1}, has two children, where the key ``string'' corresponds to the left child, and the key ``object'' corresponds to the right child. 

\begin{figure}[ht]
    \includegraphics[width=0.40\textwidth, height=0.2\textwidth]{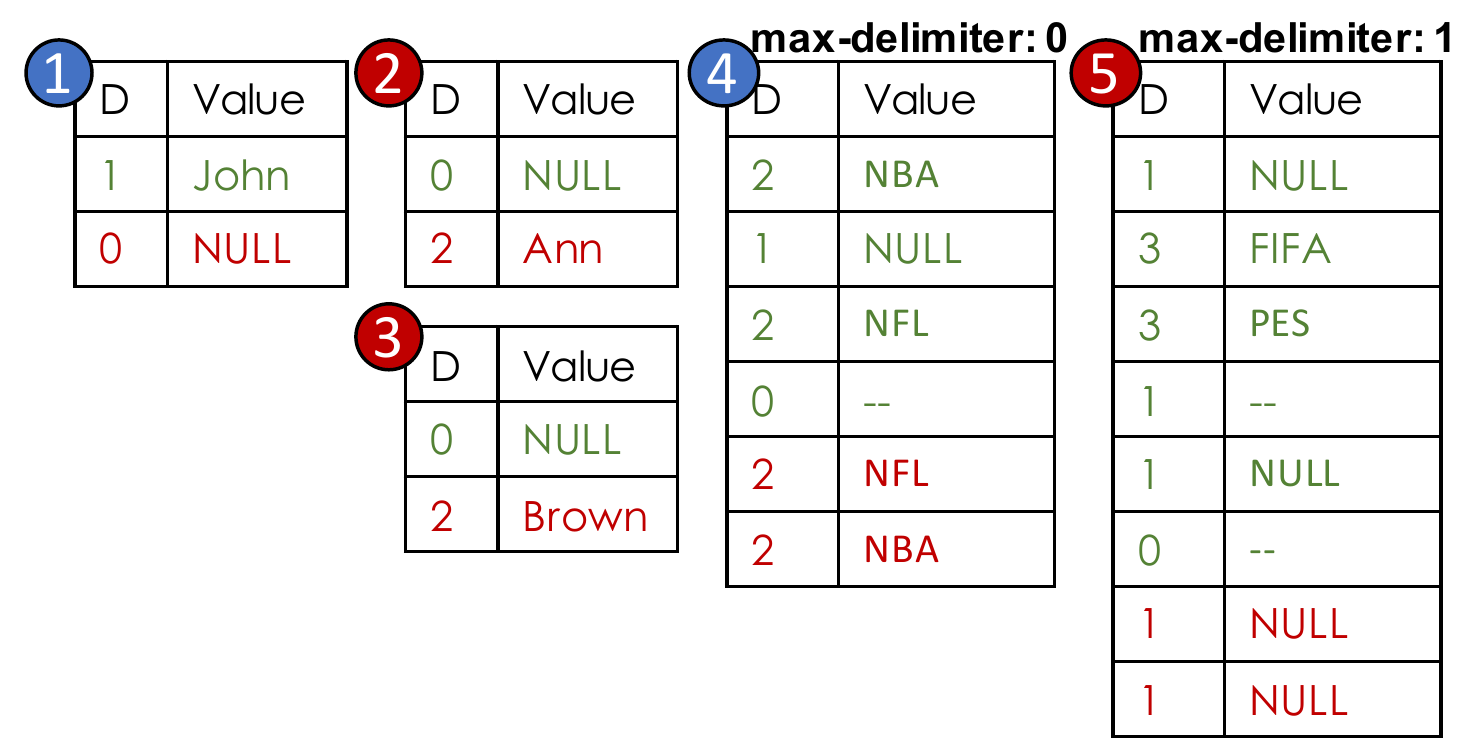}
    \vspace{-0.3cm}
    \caption{Columnar representation of the records in Figure~\ref{fig:extended-union1}}
    \label{fig:extended-union2}
\end{figure}

Based on this observation, we can columnize the unions' atomic values by treating them as object atomic values with one modification. Observe that an actual value can only be of a single type in any given record, and, hence, only a single value can be present, whereas the other atomic values associated with the union should be NULLs. To better illustrate, consider an example where the records are inserted one after another, and the schema changes accordingly. Columnizing the records' values can be performed while inferring the schema in a single pass, as in the compaction process in \cite{compact}. After inserting the first record of Figure \ref{fig:extended-union1}, we infer that field $name$ is of type string, and thus we write the string value "John" with definition level 1 as shown in column \bluecircle{1} in Figure \ref{fig:extended-union2}. In the following record, the field $name$ is an object consisting of $first$ and $last$ fields. Therefore, we change the field $name$'s type from string to a union type of string and object as was shown in Figure \ref{fig:extended-union1}. Since the second record is the first to introduce the field $name$ as an object, we can write NULLs in the newly inferred columns \redcircle{2} and \redcircle{3} for all previous records. Then, we write the values "Ann" and "Brown", with definition levels 2 in \redcircle{2} and \redcircle{3}, respectively. Recall that only a single value can be present in a union type; therefore, we write a NULL in column \bluecircle{1}. After injecting the union node in the path root $\rightarrow$ union $\rightarrow$ string, we do not change the definition level of column \bluecircle{1} from 1 to 2 for two reasons. First, union nodes are logical guides and do not appear physically in the actual records. Therefore, we can ignore the union node as being part of a path when setting the definition levels even for the two newly inferred columns \redcircle{2} and \redcircle{3}. The second reason is more technical --- changing the definition levels for all previous records is not practical, as we might need to apply the change to millions of records, were it even possible due to the immutable nature of LSM.

When accessing a value of a union type, we need to see which value is present (not NULL) by checking the values of the union type one by one. If none of the values of the union type is present, we can conclude that the requested value is NULL. In the example shown in Figure \ref{fig:extended-union1} and Figure \ref{fig:extended-union2}, accessing $name$ goes as follows. First, we inspect column \bluecircle{1}, which corresponds to the string child of the union. If we get a NULL from \bluecircle{1}, we need to proceed to the following type: an object with two fields, $first$ \redcircle{2} and $last$ \redcircle{3}. In this case, we need to inspect one of the values' definition levels, say column \redcircle{2}. If the definition level is 0, we can conclude that the value $name$ is NULL, as the string and the object values of the union are both NULLs. However, if the definition level is 1, we know that the parent object is present, but the $first$ string value is NULL. Thus, the result of accessing the field $name$ is an object in this case. Inspecting all the values of a union type is not needed when the requested path is a child of a nested type. For instance, when a user requests the value $name.last$, processing column \redcircle{3} is sufficient to determine whether the value is present or not. Thus, the results of accessing the value $name.last$ are NULL in the first record and "Brown" in the second record.

The types of repeated values (array elements) can alternate between two or more types, as in the array $games$ in Figure \ref{fig:extended-union1}. In the first record, the elements' types of the array $games$ are either a string or an array of strings. Similar to the value $name$, when accessing the value $games$, we need to inspect both columns \bluecircle{4} and \redcircle{5} to determine which element of the two types is present. When accessing the value $games$, we see that the first value's definition level in column \bluecircle{4} is 2, which is the maximum definition level of the column for the string value "NBA". In column \redcircle{5}, however, the definition level is 1, which indicates that the inner array of the union type is NULL. Thus, we know that the first element of the array $games$ corresponds to the string alternative of the union type. The following definition level 1 in column \bluecircle{4} indicates that the second element is NULL, whereas it is 3 in column \redcircle{5}, which is the maximum definition level of the column. Hence, the second element of the array is of type array of strings. The two values with definition levels 3 and the following delimiter with definition level 1 correspond to the two elements \lstinline[language=AQLSchema,basicstyle=\noindent\small]|["FIFA", "PES"]| of the first record. Following the delimiter, the definition level 1 in column \redcircle{5} indicates that the third element of the outer array is NULL. However, the definition level 2 in column \bluecircle{4} for the value "NFL" indicates that the third value of the outer array is a string. The delimiter 0 in both columns \bluecircle{4} and \redcircle{5} indicate the $games$'s end of values for the first record. The final result of accessing the value $games$ in the first record, therefore, is \lstinline[language=AQLSchema,basicstyle=\noindent\small]|["NBA", ["FIFA", "PES"], "NFL"]|, which preserves the original value of the record shown in Figure \ref{fig:extended-union1}. In the second record, we can see that the array consists of two elements, both of type string. Hence, the two NULL values in column \redcircle{5} indicate that neither of the two elements is of the array of strings alternative of the $games$'s union type.

\subsubsection{\textbf{LSM Anti-matter:}}
In Section \ref{sec:asterix}, we briefly explained the process of deleting records in an LSM-based storage engine using anti-matter entries. Anti-matter entries are special records that contain the key of the deleted record. In the example shown in Figure \ref{fig:dremel-columns}, the $id$'s maximum definition level is 0, as it is a required value. To represent such anti-matter entries in our proposed columnar format, we set the maximum definition level for the records' primary key(s) values to 1, i.e., their definition levels' values could be either 0 or 1. The definition level here does not indicate whether the value is NULL or present; instead, it indicates whether the primary key value corresponds to a record or to anti-matter. When the definition level of a primary key value is 0, it indicates that the primary key value is an anti-matter entry for a previously inserted record with the same key. When the definition level is 1, we know it is a newly inserted record.

\subsubsection{\textbf{Record Assembly:}}
When accessing a nested value such as the nested value $name$ in Figure \ref{fig:dremel-columns} in our approach, all of its atomic values (i.e., first and last) are stitched together to form an object (e.g., \lstinline[language=AQLSchema,basicstyle=\noindent\small]|{"first": "John", "last": "Smith"}|) using the same record assembly automaton used in \cite{dremel}. Also, we use the same Dremel algorithm to assemble repeated values (arrays). However, a difference is that we use delimiters to transition the state when constructing the arrays instead of the repetition levels as in Dremel.

%
%

\section{Columnar Formats in LSM Indexes}
\label{sec:store}
A major feature of representing records' values as contiguous columns, as in our extended Dremel format, is that it allows us to encode and possibly compress the values of each column according to its type to reduce the overall storage footprint. The immutability of LSM-based storage engines makes them especially good candidates for storing encoded values as in-place updates are not permitted. In this work, we propose two layouts for storing the columns in LSM-based document stores: (i) AsterixDB Partitioned Attributes Across \textit{(APAX)} and (ii) AsterixDB Mega Attributes Across \textit{(AMAX)}. We have implemented and evaluated both layouts in Apaches AsterixDB, hence the names. In the following sections, we first briefly explain the supported techniques used to encode the column values. Then, we detail the structures of both the APAX and AMAX layouts. Next, we describe the lifecycle of reading and writing the columns, and finally, we cover challenges related to answering queries with secondary indexes.

\subsection{Encoding}
Apache Parquet offers a rich set of encoding algorithms \cite{parquet-encoding} for different value types, including bit-packing, run-length encoding, delta encoding, and delta strings. In this work, we use all of Parquet's encoding algorithms except for dictionary encoding, which requires additional pages to store the dictionary entries. (We leave potential support for dictionary encoding for future work).

\subsection{\textit{APAX} Layout}
\label{sec:apax}
Ailamaki et al. proposed Partition Attributes Across (PAX)~\cite{pax}, a cache-friendly page layout as compared to the commonly used row-major layout (or N-ary Storage Model, a.k.a., slotted pages). PAX pages store each attribute's values contiguously in minipages. APAX minipages can be reached by relative pointers stored in the pages' header. Within a PAX page, fixed-length and variable-length values are stored in F-minipages and V-minipages, respectively. Along with the values, F-minipages contain a bit vector to indicate whether a value is present or NULL. The V-minipage stores values similar to the F-minipage; however, instead of the presence bits, the V-minipage uses values' offsets to determine the lengths of each variable-length value. NULL offsets (e.g., offset zero) indicate NULL values on the V-minipage.

\begin{figure}[ht]
    \centering
    \includegraphics[width=0.45\textwidth]{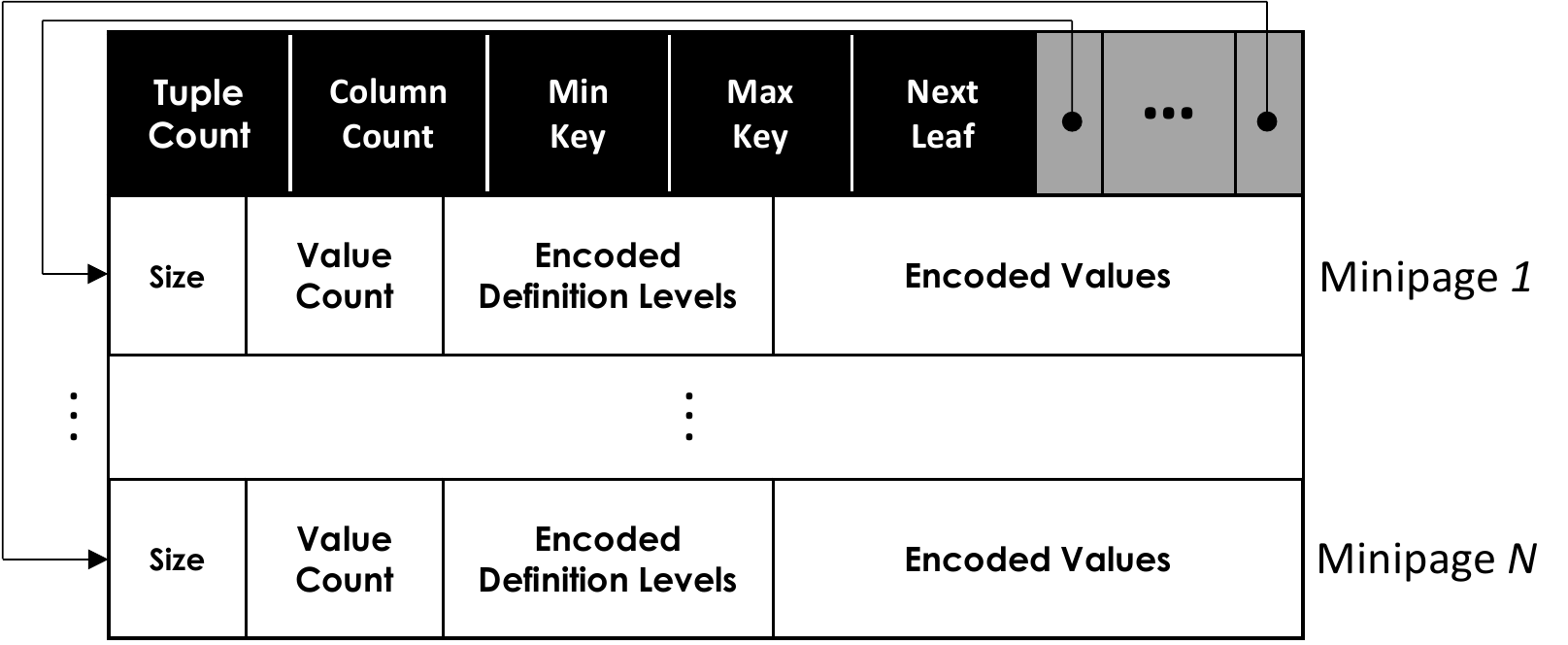}
    \caption{APAX page layout}
    \label{fig:apax}
\end{figure}

Our APAX layout is a modified version of the PAX layout, as shown in Figure \ref{fig:apax}, where fixed-length and variable-length values are encoded and stored in homogeneous mini-pages (i.e., no F-minipages and V-minipages). Thus, APAX is agnostic of its minipages' contents, and it is up to the minipages' readers and decoders to interpret the minipages' content, where the inferred schema determines the minipages' appropriate readers and decoders. Figure \ref{fig:apax} shows the organization of an APAX page. The reader will read the first four bytes to determine the size of the encoded definition level. Then, it will pass both the encoded definition levels and the encoded values to the appropriate decoders. The resulting decoded definition levels and values are then processed, as explained earlier in Section \ref{sec:extended-dremel}. As in PAX, we can reach each minipage via pointers stored in the APAX page header. Since APAX pages reside as leaf pages in a B$^+$-Tree, we store the minimum and the maximum keys (primary keys) within the APAX page header. By doing so, we can access their minimum and maximum keys directly when performing B$^+$-tree operations (e.g., search) without the need to decode the primary keys. The header also stores the number of minipages (or columns) and the number of records stored in the APAX page.

\subsection{\textit{AMAX} Layout}
\label{sec:amax}
\begin{figure}[ht]
    \centering
    \includegraphics[width=0.48\textwidth]{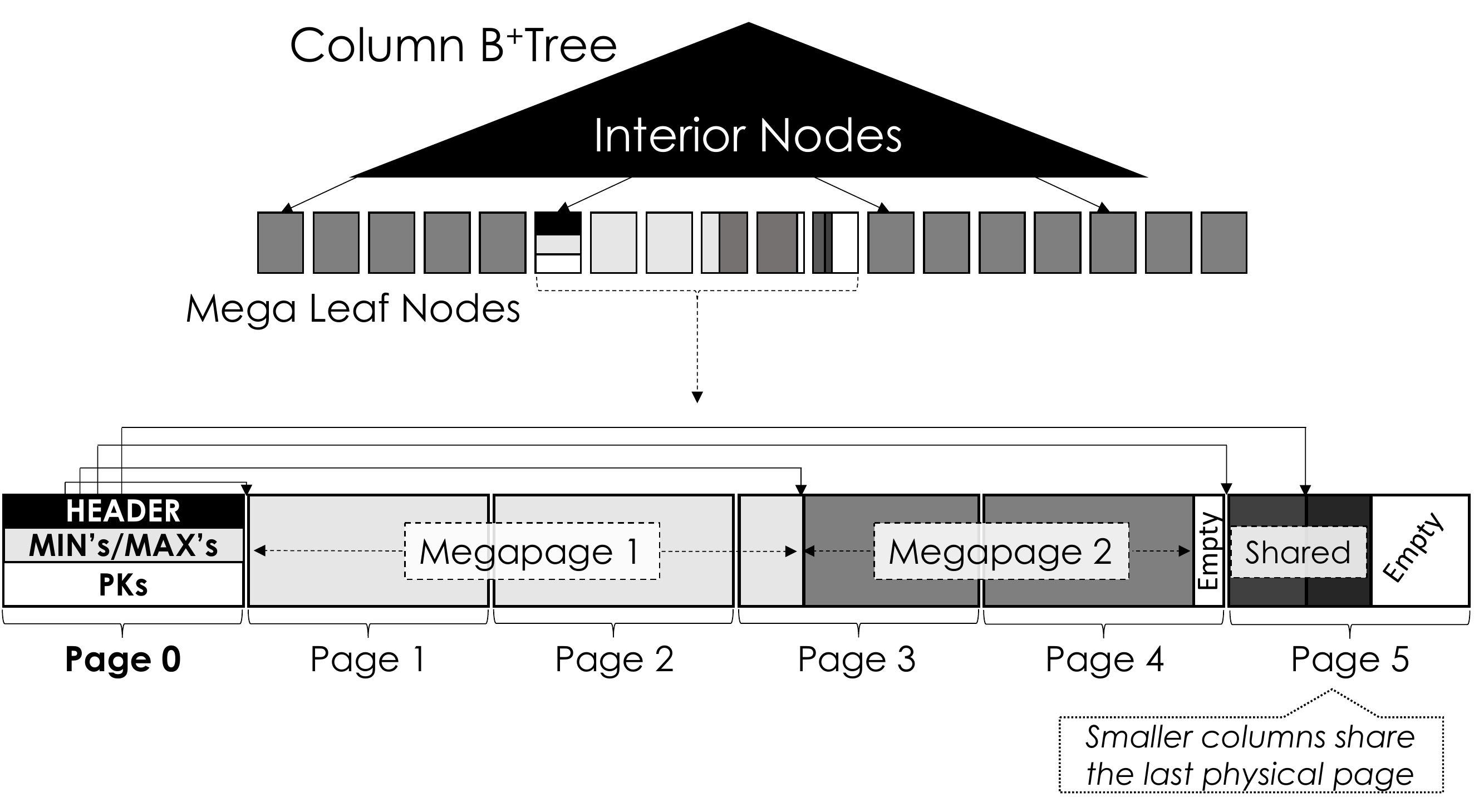}
    \caption{AMAX multi-page layout in a B-$^+$tree}
    \label{fig:amax}
\end{figure}

The PAX and APAX layouts each store different columns within a page, and hence in both layouts, we need to read the entire page, regardless of which columns are needed to answer a query. In AMAX, we stretch the minipages of APAX to become megapages that can occupy more than one physical data page. Figure \ref{fig:amax} illustrates the structure of the AMAX pages in a B-$^+$tree, where a mega leaf node consists of multiple physical pages. Each mega leaf node starts with Page 0 in the AMAX layout and consists of three segments. The first segment stores the page header, which contains the same information as in the APAX header. Following the page header, it stores fixed-length prefixes of the minimum and maximum values for each megapage (or column). Each minimum and maximum prefix pair occupy 16 bytes (or 8-byte each), and they are used to filter out AMAX pages that do not satisfy a query predicate (e.g., $age > 20$). In its last segment, Page 0 stores the encoded primary key(s) values.

Each megapage of the AMAX layout corresponds to a single column (as in APAX minipages). The megapages are ordered by their size from largest to the smallest. In other words, we store the largest megapage's physical pages contiguously first on disk, followed by the physical pages of the second-largest megapage, and so on. This ordering of megapages allows for better utilization of the empty space of the physical pages. 
For example, after writing $Megapage~1$, the physical $Page~3$ in Figure \ref{fig:amax} is mostly empty, and thus, we allow $Megapage~2$ to share the same physical $Page~3$ with $Megapage~1$. After writing $Megapage~2$, note that $Page~4$ is not full. A user-provided parameter (called the $empty-page-tolerance$) allows the AMAX page writer to tolerate a certain percentage of a physical page to be empty if the next column to be written does not fit in the given empty space. Tolerating smaller empty spaces can minimize the number of pages to be read from when retrieving a column's value. The content of the megapages is similar to APAX minipages, and we use the same readers and decoders for interpreting their content with one difference: The first physical page of each variable-length values' megapages stores the actual minimum and maximum values, as the prefixes of the minimum and maximum values on Page 0 are not decisive. The actual minimum and maximum values allow us to verify whether the values of a variable-length megapage can be within the range of the predicate.

\subsection{Reading}
\label{sec:reading}
When a user submits a query, the compiler optimizes the query and generates a job that will access the appropriate collections (or datasets in AsterixDB's terminology) and project the required attributes from the resulting records. The generated job is then distributed to all partitions in each NC for execution. Before execution, each partition consults the inferred schema to determine the columns needed (i.e., APAX minipages or AMAX megapages) for executing the query. Moreover, in AMAX, we only read the physical pages that correspond to the columns needed by the query. For each requested column, we have an iterator to go over the columns' values. If a query contains a filtering predicate (e.g., \lstinline[language=AQLSchema,basicstyle=\noindent\small]|WHERE age > 20|), we also use the prefixes of the minimum and maximum values to skip reading the entirety of the requested columns of a leaf page that do not satisfy the query predicate.

When reading from an LSM index, we need to inspect the records stored in all of the components (including the in-memory component) and merge the resulting records to reconcile any deleted or upserted records. Thus, deleted and replaced records (by upsert) are ignored and will not appear in the final result of the query. When reading APAX or AMAX pages, we need to (i) perform the same reconciliation process as in the row-major layout. Also, we need to (ii) process the in-memory component's records, which are still in a row-major layout. To address those two requirements, we implement an abstracted view of a ``tuple'', whether in row-major or column-major format, resulting from reading an LSM component. Additionally, we provide tuple comparators to compare the primary keys of two tuples resulting from two different LSM components. These comparators are agnostic of whether the tuples are from in-memory components (row-major) or on-disk components (column-major).  

Reconciling tuples in a row-major layout is performed simply by ignoring the current tuple deleted or replaced record and going to the next one using the tuple's offset stored on the slotted page. However, in a column-major layout, ignoring a tuple means ignoring its values for all the columns involved in the query by advancing the columns' iterators by one step. Doing so eagerly is inefficient, as (i) we would need to touch multiple regions of the memory, resulting in many cache misses, and (ii) we would need to decode the values each time we advance a column iterator, which could be a wasted effort as we illustrate next. Let us consider the following query:

\begin{minipage}[h]{0.1em}
\centering
\begin{center}
	
\end{center}
\begin{lstlisting}[
           language=SQL,
           showspaces=false,
           basicstyle=\sf\small,
           numberstyle=\tiny,
           commentstyle=\color{gray}
        ]
SELECT name,salary FROM Employee WHERE age>30
\end{lstlisting}
\end{minipage}

\noindent  Suppose that we have three records with primary keys 1, 2, and 5 stored in an on-disk component in a columnar layout, whether APAX or AMAX. Also, suppose the in-memory component has three records with the same primary keys, i.e., 1, 2, and 5. In this example, the records of the in-memory component will override the records of the on-disk component. If we advance every column's iterator eagerly (namely the $name$, $age$, and $salary$ columns' iterators) in order to get the next tuple, the decoding of the columns' values would be a wasted effort. For that reason, we only decode the records' primary keys during the reconciliation process, and we count the number of ignored records. Once actually accessed, we advance each column's iterator by the number of ignored records at once, ensuring that the process of advancing the iterator is performed in batches per column. Consequently, none of the columns would be decoded in our example as none were accessed.

\subsection{Writing}
\label{sec:write}
As in \cite{compact}, we exploit LSM-lifecycle events to infer the schema and split the records in row-major format into columns. During data ingestion, we first insert the records into the in-memory component in our vector-based format. Once it is full, the records of the in-memory component are flushed into a new on-disk component, during which time we infer the schema of the flushed records and split their values into columns -- storing the columns as APAX or AMAX pages. Each has different implications in terms of CPU and memory usage. In the following, we show and discuss our approach for writing the columns' values as APAX and AMAX pages.

\subsubsection{\textbf{Writing APAX Pages:}}
\label{sec:write-apax}
Determining the sizes of our APAX minipages is even more challenging than determining PAX minipages' sizes, as we incrementally encode each column's values. To address this issue, we first write the columns' values into temporary buffers, where each temporary buffer is dedicated to a single column. Once the temporary buffers have a page's-worth of values, we copy and align their contents as APAX minipages and write the resulting APAX page into the disk. We reuse the same temporary buffers to construct the following APAX pages for the remaining records of the flushed in-memory component.

\subsubsection{\textbf{Writing AMAX Pages:}}
\label{sec:write-amax}
As opposed to APAX minipages, AMAX columns can occupy one or more physical pages (megapages), while the smaller columns may share a single physical page. Initially, we do not know which columns might span into multiple physical pages, so we write the values of each column into a growable temporary buffer first. Once a temporary buffer grows beyond a configured threshold, we confiscate (or acquire) a page from the system's buffer cache, which replaces the temporary buffer for writing the columns' values. Instead of allocating a memory budget for writing columns, we use the buffer cache as a temporary buffer provider. Allocating a dedicated memory budget for writing columns might be wasteful, especially for cases where writes are not continuous (e.g., loading a dataset once and never updated). As the column size increases, we confiscate more pages from the buffer cache to accommodate the written values of that column, and those physical pages form a megapage. Once done, we write the megapages to disk largest to smallest, during which time we observe the $empty-page-tolerance$ threshold (Section \ref{sec:amax}).

Page 0 of the AMAX could also, in theory, grow to occupy multiple physical pages. However, we do not permit that, as the number of keys could grow into hundreds of thousands. Consequently, point-lookups would perform poorly, as we need to perform a linear search to find the required key; this could negatively impact both the ingestion rate and answering queries with secondary indexes, as we discuss later in Section \ref{sec:store_index}. Therefore, we limit the number of records stored in an AMAX page to 15,000 by default. The limit can be tuned for a given workload. For example, increasing the limit for scan-only workloads, where no secondary indexes are declared, would improve the query execution time while not impacting the ingestion performance.

\subsubsection{\textbf{Impact of LSM Merge Operations:}}
\label{sec:store-merge}
From time to time, an LSM merge operation is scheduled to compact the on-disk components. In both the AMAX and APAX layouts, we need to read the columns' values from different components and write them again into a newly created merge component. The order in which the columns' values are written is determined by records' keys from each component, and the columns' values that correspond to the smallest keys are written first. Similar to the issue discussed in Section \ref{sec:reading}, eagerly reading the columns' values in each component would result in touching different regions in memory, which is not cache-friendly. To remedy this issue, we employ what we call the vertical merge. In the vertical merge, we first merge the primary keys resulting from the different components, and we record the sequence of the components' IDs. Then, we merge the values of each column from the different components similarly using the order of the recorded sequence components IDs from merging the keys. This vertical merge of the columns ensures that only one column is merged at a time. Thus, the number of memory regions that we need to read from is equal to the number of merging components instead of the number of columns times the number of components. In the AMAX layout, this is important as we only need to read one megapage at a time from each component instead of all the components' megapages, which could pressure the buffer cache.

Another issue when merging the columns is the CPU cost of decoding and encoding the columns' values, especially for datasets withlarge number of columns. In our initial experiments, this CPU cost became more apparent during concurrent merges, peaking at 800\% on an 8-core machine, which could render the system unusable for users who want to query their data. The potential resource saturation resulting from concurrent merges in LSM-based storage engines is well-known \cite{concurrent_merges}, and limiting the number of concurrent merges can remedy this issue. Conseqently, we limit the number of concurrent merges for APAX and AMAX layouts by half the number of parititons by default. Limiting the number of concurrent merges may stall writes and negatively impact the ingestion rate \cite{luo2019performance}, but writing the records in a columnar format can reduce the overall storage footprint, which means less I/O. We believe an extensive evaluation, as in \cite{luo2019performance}, should be conducted to measure those tradeoffs; however, it is beyond the scope of the current paper, so we leave it for future work.

\subsection{\textbf{Point Lookups and Secondary Indexes}}
\label{sec:store_index}
In LSM-based storage engines, one can blindly insert new records into the in-memory component without checking if a record with the same key exists (to ensure the uniqueness of the primary keys), as records with identical keys are reconciled at the query time. However, this mechanism only applies to the primary index and not to its associated secondary indexes. For a secondary index, in addition to adding the new entry, we also need to clean out the old entry (if any). Thus, we need to perform a point lookup to fetch the old value from the primary index to clean the old value by adding appropriate anti-matter entries in each secondary index. Consequently, during data ingestion, point lookups are performed for each newly inserted record to check if a record with an identical key exists. If so, its old values are retrieved to maintain the secondary indexes correctness.

Performing point lookups against datasets stored in APAX or AMAX layouts is more expensive than their in row-major counterparts, as we need to decode primary keys and linearly search for the requested value in both the APAX and AMAX layouts. The number of keys on slotted pages is usually smaller than in the APAX and AMAX layouts, and here we can search for a key in a logarithmic time (assuming the records are sorted). Point lookups in the AMAX layout are even more expensive than the APAX layout since the number of keys stored on Page 0 of the AMAX layout is significantly larger. To alleviate the cost of point lookups for APAX and AMAX layouts, we use a "primary key index", a secondary index that stores only primary keys, to first see if that a record with an identical key exists \cite{luo2019efficient, luo2020lsm}. If the primary key index does not yield any keys, we skip accessing the primary index, as the newly inserted key does not correspond to an older record. Thus, we need to actually access the primary index only to update old records.

When answering queries (e.g., range queries), we first search the appropriate secondary index, which yields the primary keys of records that satisfy the query predicate. Then, we sort the resulting primary keys in ascending order. Finally, we perform a point-lookup using the sorted primary keys to retrieve the records that satisfy the query predicate. Luo et al.'s generalized approach \cite{luo2019efficient} exploits the ordered keys to perform these point lookups in batches while preserving the state of the LSM cursor to reduce the cost of subsequent point lookups. This approach allows us to read the columns' values in a single pass by accessing the values of the first record with the smallest key followed by the record with the second smallest key, etc., without the need to start over each time. However, if we were to skip sorting the primary keys resulting from accessing the secondary index, then we would need to decode the columns for each point lookup, making the use of secondary indexes, in most cases, slower than scanning the entire records.

\section{Code Generation}
\label{sec:codegen}

\begin{figure}[ht]
    \centering
    \includegraphics[width=0.35\textwidth]{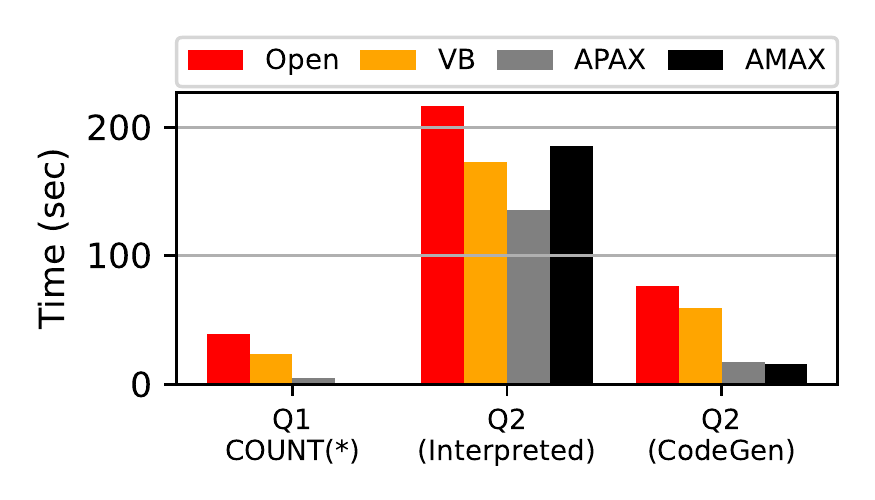}
    \vspace{-0.3cm}
    \caption{Execution time with and without code generation}
    \label{fig:codegen-expr}
\end{figure}

\begin{figure}[ht]
    \centering
    \includegraphics[width=0.45\textwidth]{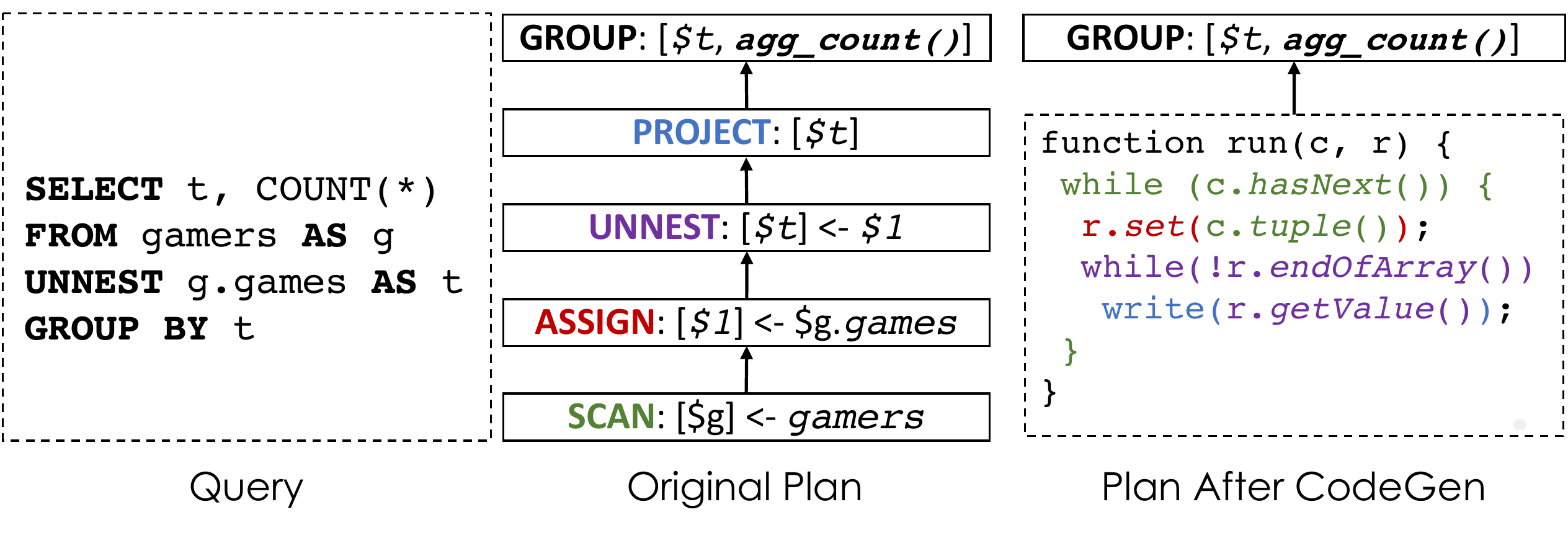}
    \caption{Code generation workflow}
    \label{fig:codegen-workflow}
\end{figure}

When we first evaluated the performance of querying records in the APAX and AMAX layouts in Apache AsterixDB, we saw that, in certain cases, their performance gains were negligible compared to the row-major formats, despite the storage savings of the columnar layouts. Figure \ref{fig:codegen-expr} shows our early evaluation results, where we ran two queries Q1 and Q2, against records in different formats: (i) AsterixDB's schemaless format (Open), (ii) the Vector-based format (VB), (iii) APAX, and (iv) AMAX. Both Open and VB are row-major formats. Q1 only counts the number of tuples, whereas Q2 is a $GROUP$ $BY$ aggregate query, similar to the query shown in Figure~\ref{fig:codegen-workflow}. The storage savings in both the APAX and AMAX layouts, as we will see later in Section~\ref{sec:expr}, significantly improved the performance of Q1. However, Q2 (Interpreted) took more time to execute against the AMAX format than the VB format. One reason for these negligible (or even negative) performance gains is the Hyracks batch-at-a-time execution model, in which the tuples are materialized between operators. However, the major reason was the cost of reassembling nested values in APAX and AMAX so that Hyracks operators can process the row-major tuples. Changing Hyrakcs to operate on columnar values natively would be a laborious task and may not yield better performance. Thus, we opted to use an approach similar to the one proposed in \cite{neumann}, where we generate a code for parts of a query plan. Figure \ref{fig:codegen-expr} shows the times to execute the same query Q2 using the code generation approach. Even though we only generate a code for parts of the plan, Q2 (CodeGen) took significantly less time to execute than Q2 (Interpreted), even for the row-major formats Open and VB. Next, we give a brief overview of the Truffle framework and how we now utilize it for code generation and query compilation. 

Due to the dynamiclly typed nature of the document data model, the generated code must handle values with heterogeneous types. Thus, we use Truffle, a framework for implementing dynamically typed languages, to produce Truffle Abstract Syntax Trees (AST) for a part of the query plan. Each node of the AST describes a language operation, such as a numerical expression (e.g., arithmetic addition) or a control flow statement (e.g., while loop statement). Also, we specify the expected behavior (or specialization as in Truffle terminology) for each expression given its inputs. For instance, the output for the logical expression $10 > "ten"$ is NULL, where NULL is the expected output when comparing two values with incompatible types in AsterixDB. After observing a few values by executing the AST in an interpreted mode, Truffle optimizes the AST and generates a bytecode to run it in the Java Virtual Machine (JVM), where the generated bytecode is optimized further to machine code.

Figure \ref{fig:codegen-workflow} shows an example of a query, its optimized plan, and the plan after replacing some of its operators with generated code. The actual generated code is a Truffle AST; however, we it show as a human-readable code (instead of an AST) for a better illustration. When a user submits a query, the query optimizer applies a set of optimization rules to the query to produce an optimized query plan. Then, we take the optimized query plan and apply an additional rule where we call two functions $produce$ and $consume$ on each operator as preseneted in \cite{neumann}. The $produce$ function asks an operator to produce its result tuples, which are then pushed to the next operator for consumption by calling the $consume$ function. Each operator in the plan calls the $produce$ function of its child (or children) recursively. Then, the resulting tuples from each child operator(s) are consumed by their parent operator. The $produce$ and $consume$ functions do not actually produce or consume any tuples but are used to translate operators into an AST. In this work, we only generate code for ``pipelining'' operators. Thus, we do not generate code for the entire query plan, but we stop the code generation process once we see a ``pipeline-breaker''. In our example, the $GROUP$ operator is a pipeline-breaker, as it requires building a hash-table for the resulting tuples (hash-group-by) or sort by them (sort-group-by) to compute the groups and their aggregate counts. 

To illustrate the code generation process, the $GROUP$ operator in Figure~\ref{fig:codegen-workflow} calls the $produce$ function of its child $PROJECT$, and the $PROJECT$ operator calls the $UNNEST$'s $produce$ function and so on until we call the $produce$ function of the $SCAN$ operator. The $consume$ function is called in the opposite direction (i.e., bottom-up). The generated code, as shown in Figure \ref{fig:codegen-workflow}, begins with the function $run$'s header, which takes two parameters $c$ and $r$, where $c$ is the tuples' \textbf{c}ursor and $r$ is the field \textbf{r}eader. The reader is pre-configured to accept a tuple as an input and produce the value of a requested field. Figure~\ref{fig:codegen-workflow} also shows which part of the plan (color-coded) produces which part of the code. First we see the generated code produced from the $SCAN$ operator, which loops through the tuples of the $gamers$ collection (or dataset). Next, the $ASSIGN$ operator contributes the code for using the reader $r$ to get the field $games$ value from the tuple. The $UNNEST$ operator contributes the $while$ loop to produce the items of the array $games$. Finally, we only write the $games$' array items, as they are the only projected values from the $PROJECT$ operator. The resulting values are then pushed to the system's regular $GROUP$ operator to compute the final result.

\section{Experiments}
\label{sec:expr}
In this section, we evaluate an implementation of the proposed techniques in Apache AsterixDB in terms of their (i) on-disk storage size after data ingestion, (ii) data ingestion rate, and (iii) performance when running analytical queries. We evaluate the performance for storing and querying records in different layouts, namely: (i) AsterixDB's schemaless record format (Open), (ii) the Vector-Based (VB) format proposed in \cite{compact}, (iii) APAX, and (iv) AMAX. Again, Open and VB are both row-major formats, whereas APAX and AMAX are columnar formats.

\noindent\textbf{Experiment Setup} We conducted our experiments using a single machine with an 8-core (Intel i9-9900K) processor and 32GB of main memory. The machine is equipped with a 1TB NVMe SSD storage device (Samsung 970 EVO) capable of delivering up to 3400 MB/s for sequential reads and 2500 MB/s for sequential writes. We used AsterixDB v9.6.0 to implement and evaluate our proposed techniques. Unless otherwise noted, we configured AsterixDB with a single NC and eight partitions (Section \ref{sec:asterix}). The eight partitions share 16GB of total allocated memory, and we allocated 10GB for the system's buffer cache and 2GB for the in-memory component budget. The remaining 4GB is allocated for use as temporary buffers for query operations such as sorting and grouping as well as transforming records into $APAX$ or $AMAX$ layouts during data ingestion. Additionally, we used 128KB for the on-disk data page size and 64KB for in-memory pages. Throughout our experiments, we used AsterixDB's page-level compression with the Snappy \cite{snappy} compression scheme to reduce the overall storage footprint.

\subsection{Datasets}
\label{sec:datasets}
In our evaluation, we used five different datasets (real, scaled, and synthetic) that differ in terms of their records' structures, sizes, and value types. Table \ref{tab:datasets-summary} lists and summarizes the characteristics of the five datasets. In Table \ref{tab:datasets-summary}, \textit{\# of Columns} refers to the number of inferred columns for records in $APAX$ or $AMAX$ layouts. 

\begin{center}
    \begin{table}[!ht]  
    \small
      \centering
      \begin{tabular}{|l|c|c|c|c|c|c|}
        \cline{2-6}
        \multicolumn{1}{c|}{} & $cell$ & $sensors$ & $tweet\_1$ & $wos$ & $tweet\_2$ \\ \hline
        Type & Real & Synthetic & Real & Real & Scaled \\ \hline
        Size (GB) & 172 & 212 & 210 & 277 & 200 \\ \hline
        \# of Records & 1.43B & 40M & 17M & 48M & 77.2M \\ \hline
        Avg. Record Size & 141B & 3.8KB & 5.3KB & 6.2KB & 2.7KB \\ \hline
        \# of Columns & 7 & 16 & 933 & 296 & 275 \\ \hline
        Dominant Type & Mix & Integer & String & String & String \\ \hline
      \end{tabular}
      \caption{Datasets summary}
      \label{tab:datasets-summary}
      \vspace{-0.9cm}
    \end{table}
    \end{center}

The $cell$ dataset (provided by a telecom company) contains information about the cellphone activities of anonymized users, such as the call duration and the cell tower used in the call. The $cell$ dataset is the only dataset we used that does not contain nested values (i.e., its data is in first-normal form or 1NF), and its scalar values' types are a mix of strings, doubles, and integers. The $sensors$ dataset contains primarily numerical values that describe the sensors' connectivity and battery statuses along with their daily captured readings. In contrast, the $wos$ dataset, as well as $tweet\_1$ and $tweet\_2$, consist mostly of string values. The $wos$ dataset, an acronym for Web of Science \footnote{We obtained the dataset from Thomson Reuters. Currently, Clarivate Analytics maintains it \cite{clarivate-wos}.} \cite{cadre}, encompasses meta-information about published scientific articles (such as authors, abstracts, and funding) from 1980 to 2014. The original dataset is in XML and we converted it to JSON using an XML-to-JSON converter \cite{xml-to-json}. After the conversion, the resulting JSON documents contain some fields (resulting from XML elements) with heterogeneous types, specifically a union of an object and an array of objects. Thus, we used the $wos$ dataset to evaluate our extensions to the Dremel format to store heterogeneous values in columnar layouts. Lastly, we obtained the $tweet\_1$ and $tweet\_2$ datasets using the Twitter API \cite{twitter-api}, where we collected the tweets in $tweet\_1$ from September 2020 to January 2021. The $tweet\_2$ dataset is a sample of tweets ($\sim$ 20GB) that we collected back in 2016, which predates Twitter's announcement of increasing the character limit from 140 to 280. We replicated the $tweet\_1$ dataset to have around 200GB worth of tweets in total. Note that the records of $tweet\_1$ and $tweet\_2$ differ in terms of their sizes and the number of columns that they have, as shown in Table \ref{tab:datasets-summary}. 

We used the $tweet\_2$ dataset for evaluating the impact of declaring secondary indexes for an update-intensive workload as we detail later in Section~\ref{sec:store_index}. Additionally, we evaluated the impact of answering queries using the created secondary indexes. We created two indexes in this experiment. The first index is on the tweet $timestamp$ values, a set of synthetic and monotonically-increasing values that mimics the time when users posted their tweets. We also created a primary key index to reduce the cost of point lookups, as discussed in Section \ref{sec:store_index}. We chose $tweet\_2$ for this experiment, instead of $tweet\_1$, since it has a moderate number of columns, which directly impacts the ingestion performance, as we discuss later in Section \ref{sec:ingestion}.

\subsection{Storage Size}
\label{sec:storage}

\begin{figure}[t]
    \centering
    \begin{subfigure}[b]{0.25\textwidth}
        \includegraphics[width=\textwidth]{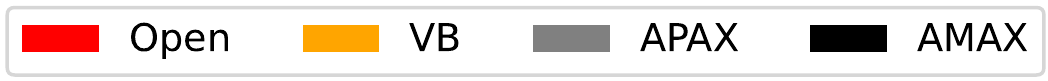}
        \vspace{-1.3em}
    \end{subfigure}
    \begin{subfigure}[b]{0.45\textwidth}
        \includegraphics[width=\textwidth, height=0.25\textwidth]{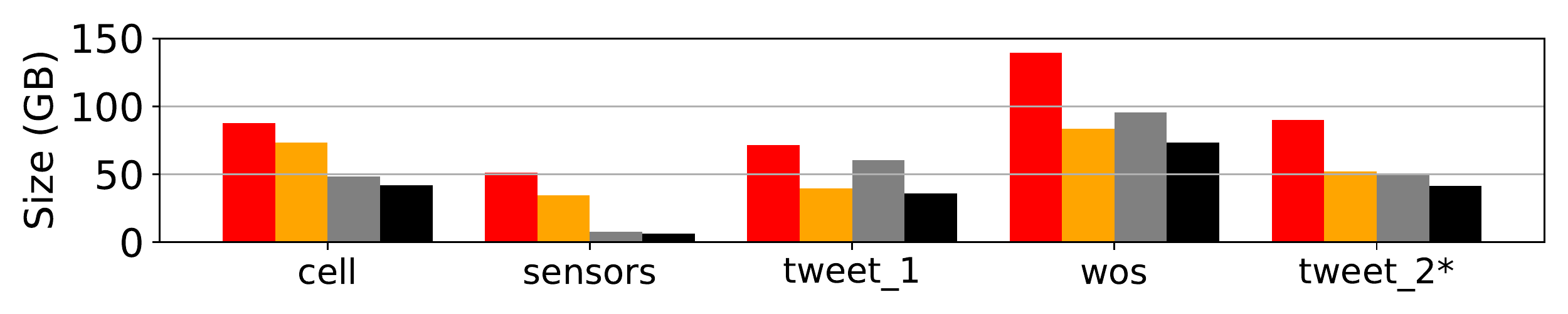}
        \vspace{-0.5cm}
        \caption{Storage size}
        \label{fig:storage}	
    \end{subfigure}
    \caption{Storage size and ingestion time --- $tweet\_2^*$ includes secondary indexes}
\end{figure}

In this experiment, we first evaluated the on-disk storage size after ingesting the five datasets: $cell$, $sensors$, $tweet\_1$, $wos$, and $tweet\_2$. Figure \ref{fig:storage} shows the total on-disk size after ingesting the five datasets using the four layouts: $Open$, $VB$, $APAX$, and $AMAX$. For the $tweet\_2$ dataset, the presented total size includes the sizes for storing the two declared secondary indexes (namely the $timestamp$ index and the primary key index).

In the $cell$ dataset, which is the only dataset in 1NF, Figure \ref{fig:storage} shows that the records in the two row-major layouts $Open$ and $VB$ took roughly the same space; the $VB$ layout took slightly less space ($\sim$ 17\% smaller) due to compaction \cite{compact}. Similarly, the records in both of the columnar layouts, $APAX$ and $AMAX$, took about the same space; however, compared to the records in the $Open$ format, the sizes are 45\% and 50\% smaller, for the $APAX$ and $AMAX$ layouts, respectively. The storage overhead reductions in the $APAX$ and $AMAX$ layouts are due to (i) storing no additional information (e.g., field names) with the values, compared to $Open$, and (ii) the values being encoded, which is not possible in the row-major layouts.

The impact of encoding in both the $APAX$ and $AMAX$ layouts becomes apparent for storing the $sensors$ dataset, where the values' are primarily numeric. Figure \ref{fig:storage} shows that the $sensors$ records in the $Open$ and $VB$ layouts took 7.2X and 4.8X more space compared to the records in the $APAX$ layout, respectively, and 8.5X and 5.6X more space compared to the records in the $AMAX$ layout, respectively. This clearly shows that the encoding of numerical values in the same domain is superior to page-level compression alone, making the columnar layout more suitable for numerical data.

In contrast to the $sensors$ dataset, the Twitter dataset $tweet\_1$ contains more textual values than numerical ones and the encoding becomes less effective. Storing this data in a columnar layout did not show a significant improvement, as shown in Figure \ref{fig:storage}, compared to the records in a row-major layout. In fact, the records in the $APAX$ layout took 35\% more space than the records in the $VB$ layout. The reason behind the high storage overhead in $APAX$ is the excessive number of columns of the $tweet\_1$ dataset, as shown in Table \ref{tab:datasets-summary}, so each minipage stores a small number of values compared to the $cell$ and $sensors$ datasets. Thus, in some instances, the encoding imposes a negative impact, as the encoded values store additional information for decoding, and this information occupies more space than the encoding saves. However, the $AMAX$ layout is not as sensitive to the number of columns, as a column can span multiple pages, and hence, the number of values is sufficient for the encoding to be effective. The storage saving from the $AMAX$ layout is negligible compared to the $VB$ layout, as encoding large textual values is relatively less effective compared to numerical values.

Storing the $wos$ dataset using the four layouts shows a similar trend, shown in Figure \ref{fig:storage}, as in the $tweet\_1$ dataset, even though the number of columns in the $wos$ dataset is not as excessive, as shown in Table \ref{tab:datasets-summary}. However, the average size of a record in the $wos$ dataset is larger than the average record size in the $tweet_1$ dataset. The reason is that some of the values in the $wos$ dataset are relatively larger than the $tweet\_1$ values. For example, the abstract text of a publication could consist of multiple paragraphs --- exceeding the number of characters of a tweet. Hence, the larger values of the $wos$ dataset limited the number of values we could store in an $APAX$ page, which again reduced the effectiveness of encoding. The $wos$ records took more space in the $Open$ layout than other layouts due to the $Open$ layout's recursive structure (as detailed in \cite{compact}), where deeply nested values require 4-byte relative pointers for each nesting level. Additionally, the $Open$ layout records embed the field names for each value, which takes more space than the other layouts.

For the last dataset, $tweet\_2$, the total storage size includes the sizes of the two declared indexes. Secondary indexes are agnostic of the records' layout in the primary index and their sizes are the same for all four layouts. Hence, the differences between the sizes for the different layouts, shown in Figure \ref{fig:storage}, correspond to the layouts' characteristics. For instance, the sizes of the records in the $VB$, $APAX$, and $AMAX$ layouts are comparable, with $AMAX$ being slightly smaller. However, the $Open$ layout took more space due to the reasons explained earlier.

\subsection{Ingestion Performance}
\label{sec:ingestion}
We next evaluated the ingestion performance for the four different layouts using AsterixDB's data feeds.  We first evaluated the insert-only ingestion performance of the $cell$, $sensors$, $tweet\_1$, and $wos$ datasets without updates. In the second experiment, we evaluated the ingestion performance of an update-intensive workload with secondary indexes using the $tweet\_2$ dataset. The latter experiment focuses on measuring the impact of the point lookups needed to maintain the correctness of the declared secondary indexes.

We configured AsterixDB to use a tiering merge policy with a size ratio of 1.2 throughout the experiments. This policy merges a sequence of components when the total size of the younger components is 1.2 times larger than that of the oldest component in the sequence. To measure the ingestion rate accurately, we used the fair merge scheduler as recommended in \cite{luo2019performance}, where the components are merged on a first-come, first-served basis. We set the maximum tolerable number of components to 5, after which a merge operation is triggered. We limited the number of concurrent merges to reduce CPU and memory consumption (Section \ref{sec:store-merge}) while merging the $APAX$ and $AMAX$ components. Additionally, we limited the number of primary keys in AMAX's Page 0 to 15,000 (default value) for the reasons discussed in Section \ref{sec:write-amax}.

\begin{figure}[t]
    \centering
    \begin{subfigure}[b]{0.25\textwidth}
        \includegraphics[width=\textwidth]{figures/exprs/legend.pdf}
        \vspace{-1.3em}
    \end{subfigure}
    \begin{subfigure}[b]{0.45\textwidth}
        \includegraphics[width=\textwidth, height=0.25\textwidth]{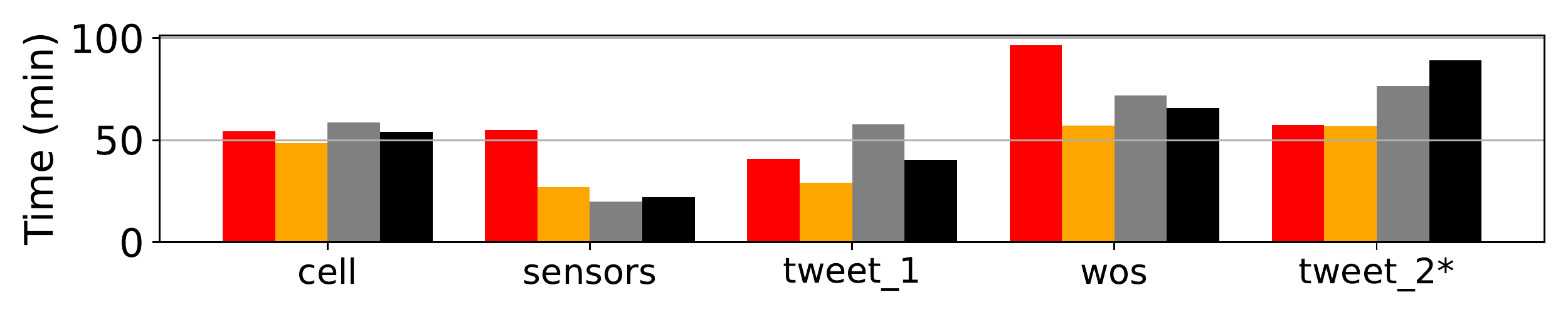}
        \vspace{-0.5cm}
        \caption{Ingestion time}
        \label{fig:ingestion}	
    \end{subfigure}
    \caption{Storage size and ingestion time --- $tweet\_2^*$ includes secondary indexes}
\end{figure}

\subsubsection{\textbf{Insert-only:}} The $cell$ dataset is the smallest in terms of the average record size and the dataset's overall size, as shown in Table \ref{tab:datasets-summary}. However, it also has the most records. In AsterixDB's original configuration (i.e., a single NC with eight partitions), the ingestion rate of the $cell$ dataset was the slowest among the datasets --- it took more than 8000 seconds to ingest the dataset under the four layouts. The main reason is that the eight partitions share the same resources, including the transaction log buffer where each partition writes entries to commit their transactions. With the $cell$ dataset's high record cardinality, writing to the transaction log buffer became a bottleneck. To alleviate the contention on the transaction log buffer, we reconfigured AsterixDB to have four NCs \footnote{Usually, only a single NC is configured per computing node.}, each with two partitions. We divided the memory budget equally among the four NCs. Using this configuration, Figure \ref{fig:ingestion} shows the time it took to ingest the $cell$ dataset using the four layouts. We see that the ingestion rate is about the same for the four layouts, as writing to the transaction log buffer is still a major bottleneck. However, the ingestion rate using the new configuration improved significantly --- from more than 8000 seconds to the vicinity of 3000 seconds.

In the $sensors$ dataset, in contrast to the $cell$ dataset, the ingestion rate varied among the different layouts as shown in Figure \ref{fig:ingestion}. Ingesting $Open$ records took more time than records in the other layouts due to the record construction cost of the $Open$ layout \cite{compact}. The recursive nature of the $Open$ layout requires copying the child's values to the parent from the leaf to the root of the record, which means multiple memory copy operations for the same value. In contrast, constructing records in the $VB$ layout is more efficient, as the values are written only once \cite{compact}. Thus, ingesting records in the $VB$ layout took 50\% less time in comparison. Recall that the records of the in-memory components are in the $VB$ format for $APAX$ and $AMAX$ (as discussed in Section \ref{sec:write}), and during the flush operation, the records are transformed into a columnar layout. Thus, the lower construction cost of the $VB$ records contributed to the higher ingestion rate of both the $APAX$ and $AMAX$ layouts. We also observed that the cost of transforming the records into a column-major layout during the flush operation and the impact of decoding and encoding the values during the merge operation were negligible.

For the $tweet\_1$ and $wos$ datasets, the cost of transforming the records into columns became more apparent due to the higher number of columns in those two datasets. Figure \ref{fig:ingestion} shows that the ingestion time for $tweet\_1$ using the $APAX$ layout was the longest. As explained earlier in Section \ref{sec:storage}, a higher number of columns can negatively affect the number of values that we can store in $APAX$ pages, and thus, more pages are required to store the ingested records. Also, recall that the columns' values are first written into temporary buffers and then copied to form an $APAX$ page (Section \ref{sec:write-apax}). Consequently, we need to iterate over the temporary column buffers (933 in total as shown in Table \ref{tab:datasets-summary}) to construct each $APAX$ page. The cost of constructing a large number of $APAX$ pages took most of the time to ingest the $tweet\_1$ dataset. We did not observe a similar behavior when constructing the $AMAX$ pages; most of the time here was spent performing LSM merges, as we need to fetch all columns for each merge operation. The ingestion performance using the $AMAX$ layout was similar to the row-major layout ($Open$) and only 25\% slower than the $VB$ layout.

The $wos$ dataset is less extreme in terms of the number of columns compared to the $tweet\_1$ dataset; however, its data contains large textual values (e.g., abstracts). As in the $sensors$ dataset, the lower per-record construction cost of the $VB$ layout was the main contributor to the performance gains (shown in Figure \ref{fig:ingestion}) for the $APAX$ and $AMAX$ layouts. Additionally, the records in the $Open$ layout took more space to store, which means that the I/O cost of the LSM flush and merge operations was higher compared to the other layouts. The ingestion performance of the $APAX$ and $AMAX$ layouts was comparable and slightly slower than the $VB$ layout, as the cost of transforming the ingested records into a column-major layout during the flush operation and decoding and encoding the values during the merge operation was higher for the $wos$ dataset compared to the $sensors$ dataset. 

\subsubsection{\textbf{Update-intensive:}} We evaluated the ingestion performance for insert-only workloads using different datasets, and we saw that the ingestion rate using columnar layouts, in general, was faster or comparable to the row-major layout $Open$. We now discuss the performance for an update-intensive workload with secondary indexes using the dataset $tweet\_2$. In this experiment, we randomly updated 50\% of the previously ingested records. The updates followed a uniform distribution where all records are updated equally. Prior to starting the data ingestion, we created two indexes: one on the primary keys, which we call a primary key index, to minimize the cost of point lookups of non-existent (new) keys. The second index is on the $timestamp$ values. Figure \ref{fig:ingestion} shows $tweet\_2$ the ingestion time for the four different layouts. The ingestion times for records in the $APAX$ and $AMAX$ were $\sim$ 24\% and $\sim$ 35\% slower than the $Open$ layout, respectively. Updating a record requires accessing the primary index to fetch the old $timestamp$ value to delete it from the $timestamp$ secondary index before inserting the updated value. Recall that the cost of searching for a value in a columnar layout is linear (v.s. logarithmic in a row-major layout), and the values need to be decoded before performing the search. Thus, with 50\% of the records being updated, the cost of updating old $timestamp$ values for the columnar layouts became higher than for the row-major layouts. Even though we only need to read the pages corresponding to the $timestamp$ in the $AMAX$ layout (i.e., less I/O cost), its update cost was higher than the $APAX$ layout. This was due to decoding large numbers of $timestamp$ values stored in $AMAX$ megapages (a CPU cost) for each update. We will soon (Section \ref{sec:query-index}) discuss the benefit of secondary indexes when answering queries; however, the cost of maintaining the correctness of secondary indexes is high for columnar layouts. Thus, one should consider how often the index would be utilized.

\begin{figure*}[t]
    \centering
    \begin{subfigure}[b]{0.3\textwidth}
        \includegraphics[width=\textwidth]{figures/exprs/legend.pdf}
        \vspace{-1.3em}
    \end{subfigure}
    \\
    \begin{subfigure}[b]{0.24\textwidth}
        \includegraphics[width=\textwidth, height=0.65\textwidth]{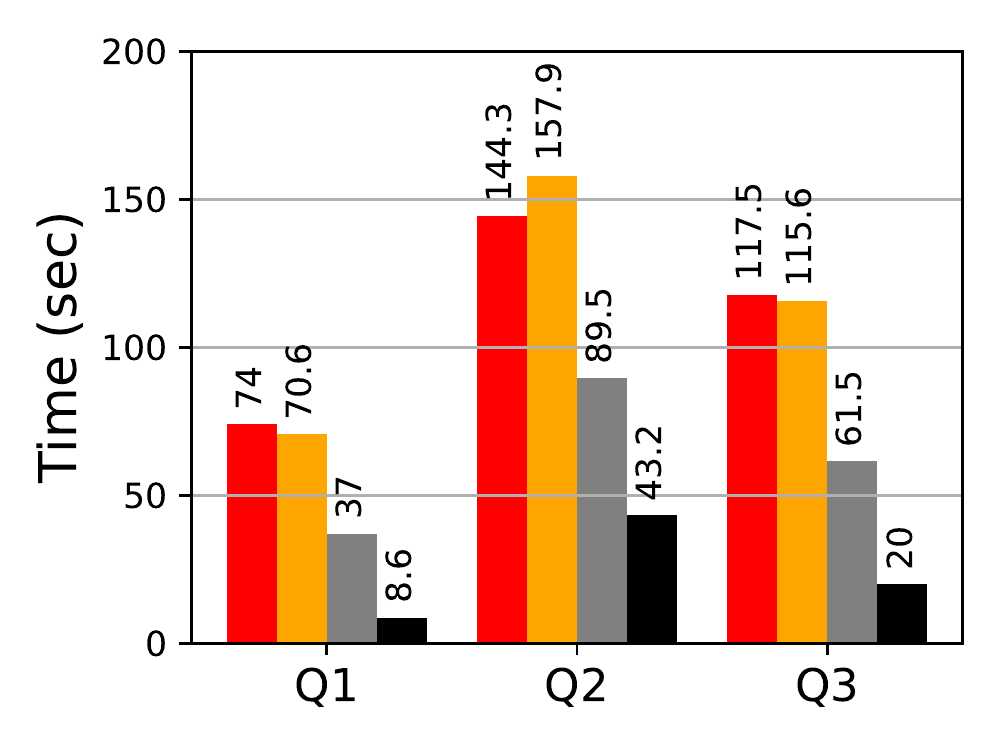}
        \caption{Query: $cell$}
        \label{fig:cdr}	
    \end{subfigure}
    \begin{subfigure}[b]{0.24\textwidth}
        \includegraphics[width=\textwidth, height=0.65\textwidth]{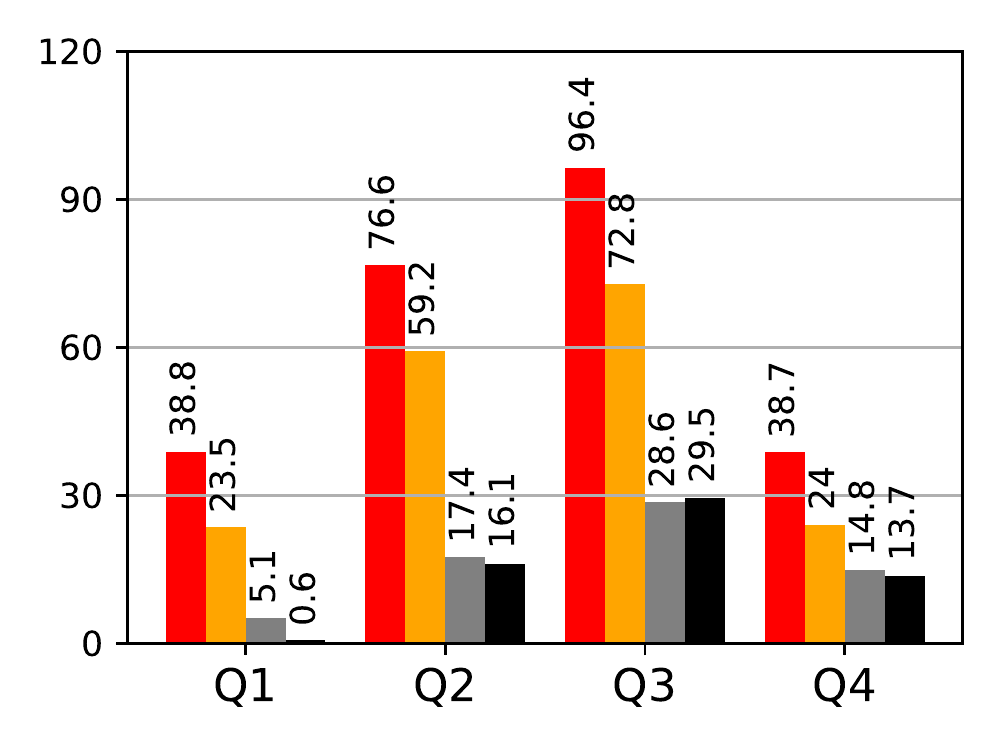}
        \caption{Query: $sensors$}
        \label{fig:sensors}	
    \end{subfigure}
    \begin{subfigure}[b]{0.24\textwidth}
        \includegraphics[width=\textwidth, height=0.65\textwidth]{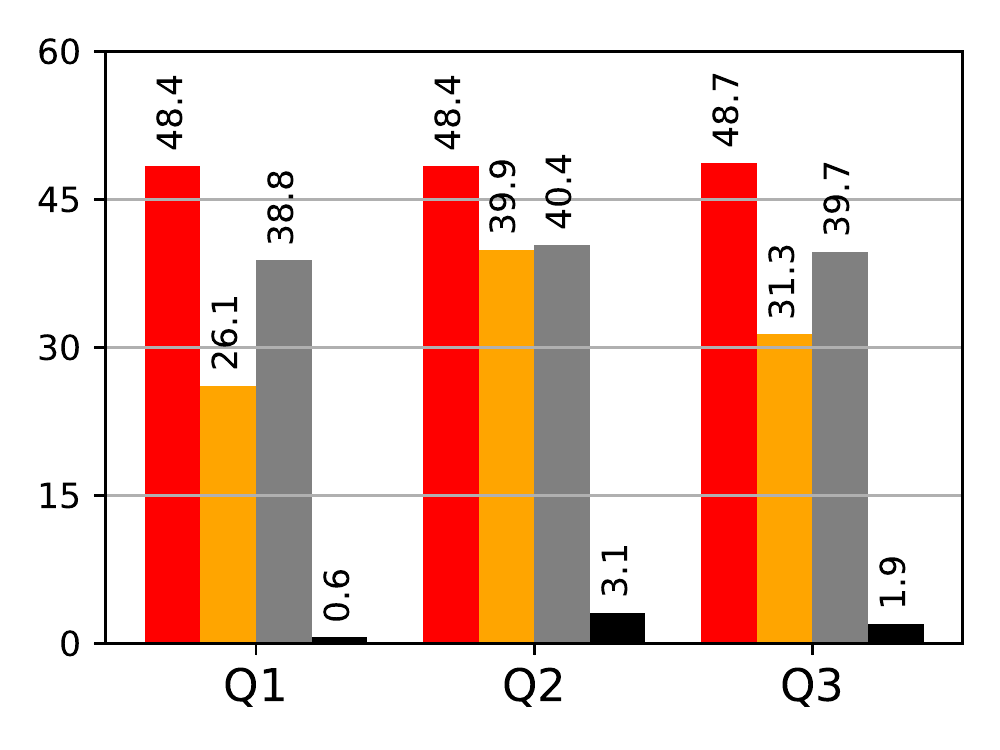}
        \caption{Query: $tweet\_1$}
        \label{fig:tweet_1}	
    \end{subfigure}
    \begin{subfigure}[b]{0.24\textwidth}
        \includegraphics[width=\textwidth, height=0.65\textwidth]{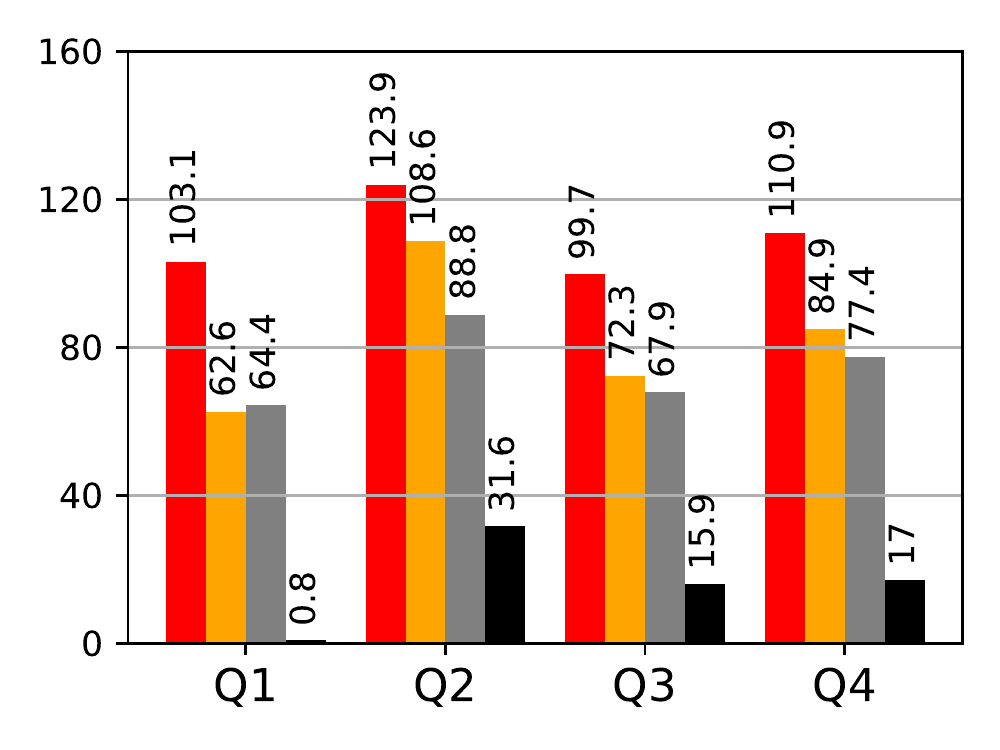}
        \caption{Query: $wos$}
        \label{fig:wos}	
    \end{subfigure}
    \caption{Experimental results}
    \label{fig:expr}
\end{figure*}
\subsection{Query Performance}
\label{sec:query}

Next, we evaluated the performance of executing different analytical queries against the ingested datasets. We first evaluated scan queries (i.e., without secondary indexes) against the $cell$, $sensors$, $tweet\_1$, and $wos$ datasets. Table \ref{tab:queries-summary} summarizes the queries used for each dataset, and the full queries are listed in \ref{appen:queries}. Q1, which counts the number of records -- \lstinline[language=AQLSchema,basicstyle=\noindent\small]|SELECT COUNT(*)|, is executed against all four datasets to measure the I/O cost of scanning records in the different layouts. We executed each query six times, and we report the average execution time for the last five. In this experiment, we report the execution times using our code generation framework (Section \ref{sec:codegen}) for the four layouts (i.e., $Open$, $VB$, $APAX$, and $AMAX$), as executing queries using the code generation technique was faster compared to AstreixDB's interpreted execution model. In the next experiment, we evaluated the performance of different queries against the $tweet\_2$ dataset using the created secondary indexes. Throughout our experiments, we evaluated the impact of accessing a different number of columns on the $AMAX$ and $APAX$ layouts for scan and index based queries.

\begin{center}
    \begin{table}[H]
        \centering
        \begin{tabular}{|c|c|l|}
        \hline
            * & Q1 & The number records\\
        \hline
        \hline
            \parbox[t]{2.5mm}{\multirow{2}{*}{\rotatebox[origin=c]{90}{$cell$}}} 
            & Q2 & The top 10 callers with the longest call durations\\
            \cline{2-3}
            & Q3 & The number of calls with durations $\geq$ 600 seconds\\
        \hline
        \hline
            \parbox[t]{1.5mm}{\multirow{3}{*}{\rotatebox[origin=c]{90}{$sensors$}}} 
            & Q2 & The maximum reading ever recorded\\
            \cline{2-3}
            & Q3 & The IDs of top 10 sensors with maximum readings\\
            \cline{2-3}
            & Q4 & Similar to Q3, but for readings in a given a day\\
        \hline
        \hline
            \parbox[t]{2mm}{\multirow{3}{*}{\rotatebox[origin=c]{90}{$tweet\_1$}}} 
            & Q2 & The top 10 users who posted the longest tweets \\
            \cline{2-3}
            & Q3 & The top 10 users with highest number of tweets that\\
            &    & contain a popular hashtag\\
        \hline
        \hline
            \parbox[t]{2mm}{\multirow{6}{*}{\rotatebox[origin=c]{90}{$wos$}}} 
            & Q2 & The top 10 scientific fields with the highest number\\
            &    & of publications \\
            \cline{2-3}
            & Q3 & The top ten countries that co-published the most\\
            &    & with US-based institutes\\
            \cline{2-3}
            & Q4 & The top ten pairs of countries with the largest\\
            &    & number of co-published articles\\
        \hline
        \end{tabular}
        \caption{A summary of the queries used in the evaluation}
        \label{tab:queries-summary}
        \vspace{-4em}
    \end{table}
\end{center}

\subsubsection{\textbf{\textit{cell} Dataset:}}
Figure \ref{fig:cdr} shows the execution time for the three queries (summarized in Table \ref{tab:queries-summary}) executed against the $cell$ dataset using the four different layouts. The execution times for Q1 against different layouts correlated with their storage sizes shown in Figure \ref{fig:storage}, except for the $AMAX$ layout, where Q1 took the least time to execute --- about 88\% faster than the $Open$ and $VB$ layouts. As Q1 only counts the number of records, we only need to count the number of primary keys on Page 0 of the $AMAX$ layout --- thereby minimizing the I/O cost. This is also true for Q1 against the other datasets, as discussed in the following sections. In contrast to Q1, Q2 requires grouping, aggregating, and sorting to compute the query's results, and hence, it takes more time to execute. For the $AMAX$ layout, in addition to the primary keys on Page 0, Q2 accesses two more columns (the caller ID and the call duration columns), which means that more pages were accessed to execute Q2. Despite the additional costs, the execution times for Q2 showed a similar trend as in Q1. Querying the $APAX$ and $AMAX$ formats were 38\% and 70\% faster than the $Open$ layout, and 40\% and 72\% than the $VB$ layout, respectively. The slowdown of querying $VB$ is due to how the fields' values are lineraly accessed, as detailed in \cite{compact}. Q3 also shows a similar trend as Q1, where the I/O costs of the two columnar layouts were the smallest.

\subsubsection{\textbf{\textit{sensors} Dataset:}}
Querying the $sensors$ dataset shows similar trends as in the $cell$ dataset, where the queries' execution times (Figure \ref{fig:sensors}) correlated with the formats' storage sizes (Figure \ref{fig:storage}). Executing Q1 against the $AMAX$ layout took 0.65 seconds, and it took 5.1 seconds for $APAX$. As in the $cell$ dataset, Q1 only read Page 0 of $AMAX$, and hence, it was the fastest. For Q2 -- Q4 (summarized in Table \ref{tab:queries-summary}), the execution times for the $APAX$ and $AMAX$ records were comparable, as they took 7.7GB and 6.5GB to store the data, respectively, which is less than the 10GB of memory allocated for the system's buffer cache. Thus, AsterixDB was able to cache the $APAX$ and $AMAX$ records in memory and eliminate the I/O cost. For the two row-major layouts, it was faster to execute the queries against $VB$ than $Open$, as $VB$ took less space.

\subsubsection{\textbf{\textit{tweet\_1} Dataset:}} For $tweet\_1$'s queries (Table \ref{tab:queries-summary}), we observed an order of magnitude improvement in the query performance using the $AMAX$ layout, vs. the other layouts. $VB$ and $AMAX$ used comparable space to store the $tweet\_1$ data; however, reading only the columns involved in the queries for the $AMAX$ layout improved their execution times significantly. For example, Q1 took only 0.6 seconds to execute against the $AMAX$ format compared to 48.4, 26.1, and 38.8 seconds for $Open$, $VB$, and $APAX$, respectively. For Q2, $AMAX$ took 3.1 seconds to execute vs. 48.5, 39.9, and 40.3 seconds for $Open$, $VB$, and $APAX$, respectively. Storing and querying the $tweet\_1$ dataset using $APAX$ showed less improvement than for the $sensors$ dataset. Excluding the $AMAX$ layout, the $VB$ layout, in comparison, was more suitable for storing and querying a text-heavy Twitter dataset, as its records took less space to store and less time to query.

\subsubsection{\textbf{\textit{wos} Dataset:}} The $wos$ dataset is the last one used to evaluate scan-based queries. As mentioned earlier in Section \ref{sec:datasets}, the $wos$ dataset contains several values with heterogeneous types. We used this dataset to evaluate the impact of querying over heterogeneous types for the columnar layouts. Specifically, Q3 and Q4 (Table \ref{tab:queries-summary}) access the authors' affiliated countries, which is stored as either an array, for articles with multiple co-authors, or as an object, for single-authored articles. Figure \ref{fig:wos} shows the execution times for Q1 - Q4, where $AMAX$ was the fastest to query. Q1 took only 0.83 seconds to execute, compared to 103.1, 62.5, and 64.4 seconds for $Open$, $VB$, and $APAX$, respectively. For Q2 - Q4, $AMAX$ improved their execution times by at least 64\% compared to the other layouts. The queries' execution times against $APAX$ were slightly shorter than the $VB$ layout. Thus, both the $APAX$ and $AMAX$ layouts can efficiently handle values with heterogeneous types, and the impact of mixed types on query performance was negligible.

\begin{figure}[h]
    \centering
    \begin{subfigure}[b]{0.3\textwidth}
        \includegraphics[width=\textwidth]{figures/exprs/legend.pdf}
        \vspace{-1.3em}
    \end{subfigure}
    \begin{subfigure}[b]{0.23\textwidth}
        \includegraphics[width=\textwidth]{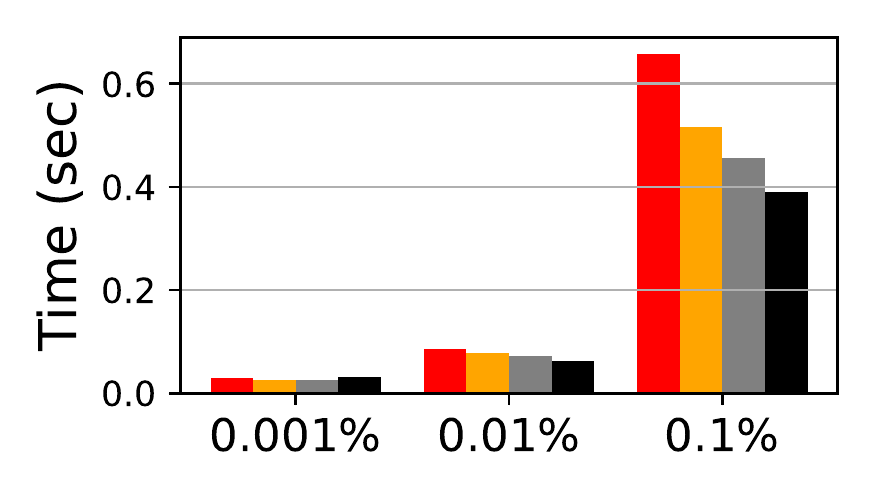}
        \caption{Low selectivity}
        \label{fig:secondary_low}
    \end{subfigure}
    \begin{subfigure}[b]{0.23\textwidth}
        \includegraphics[width=\textwidth]{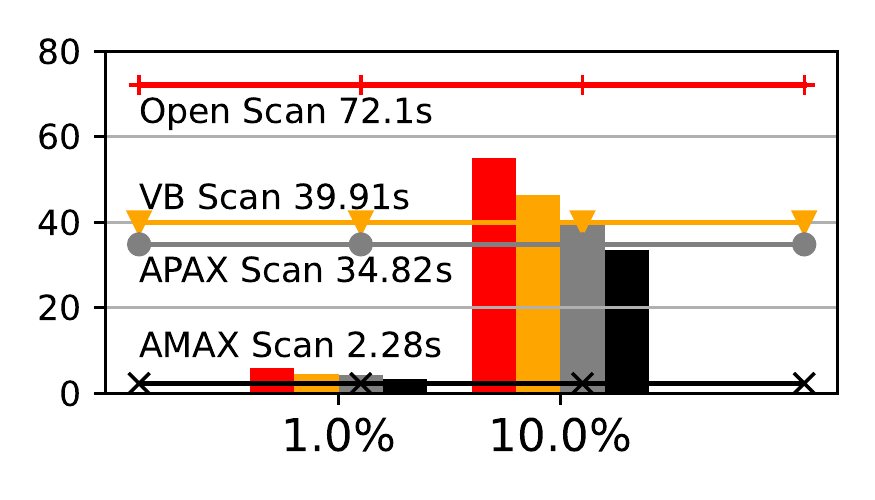}
        \caption{High selectivity}
        \label{fig:secondary_high}
    \end{subfigure}
    \vspace{-0.2cm}
    \caption{Query with secondary index}
    \label{fig:secondary-count_star}
    \vspace{-0.35cm}
\end{figure}

\begin{figure}[h]
    \centering
    \begin{subfigure}[b]{0.25\textwidth}
        \includegraphics[width=\textwidth]{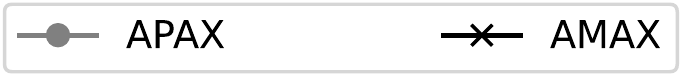}
        \vspace{-1.3em}
    \end{subfigure}
    \begin{subfigure}[b]{0.4\textwidth}
        \includegraphics[width=\textwidth,height=0.36\textwidth]{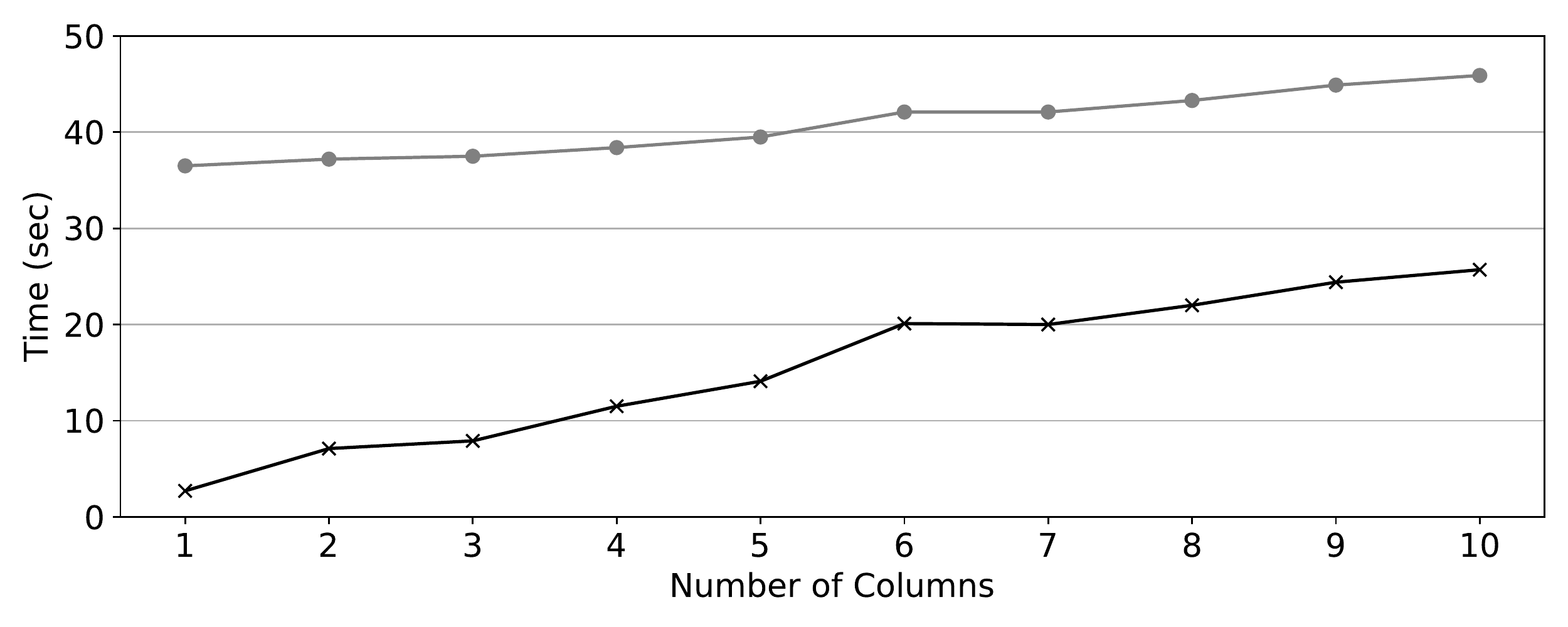}
        \caption{Execution Time}
        \label{fig:colunm_time}
    \end{subfigure}
    \begin{subfigure}[b]{0.164\textwidth}
        \includegraphics[width=\textwidth]{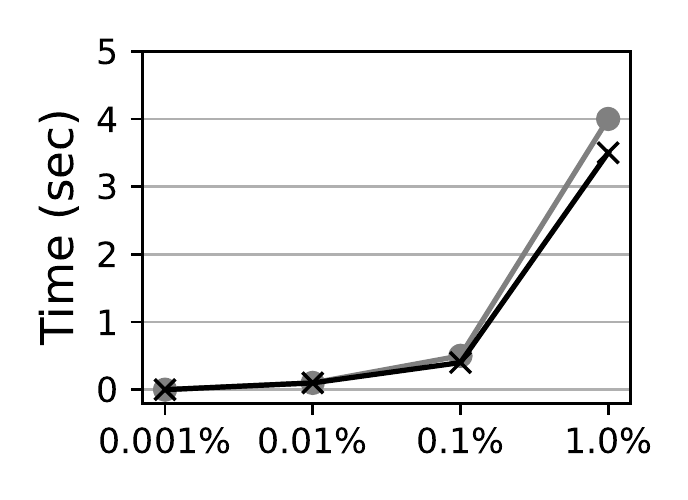}
        \caption{1 Column}
        \label{fig:column_1}
    \end{subfigure}
    \begin{subfigure}[b]{0.150\textwidth}
        \includegraphics[width=\textwidth]{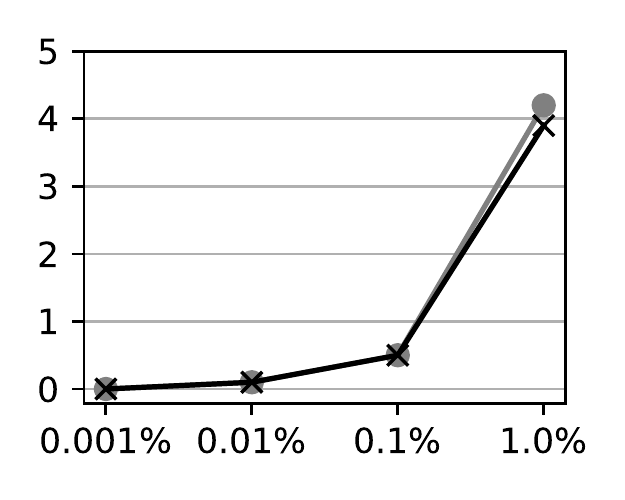}
        \caption{2 Columns}
        \label{fig:column_5}
    \end{subfigure}
    \begin{subfigure}[b]{0.150\textwidth}
        \includegraphics[width=\textwidth]{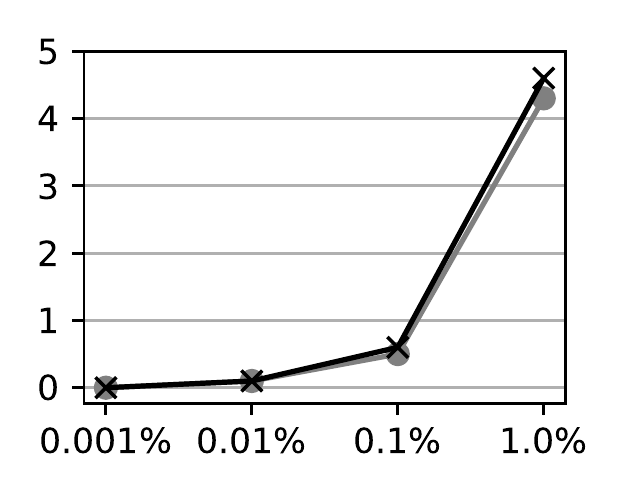}
        \caption{10 Columns}
        \label{fig:column_10}
    \end{subfigure}
    \vspace{-0.2cm}
    \caption{Impact of accessing different number of columns: (a) and (b) for scan-based, and (c) - (e) for index-based queries}
    \label{fig:column-count_star} 
    \vspace{-0.35cm}
\end{figure}

\subsubsection{\textbf{\textit{tweet\_2} Dataset:}} \label{sec:query-index} We used the $tweet\_2$ dataset to evaluate the impact of secondary indexes on query performance for the four different layouts. We used a created $timestamp$ secondary index to run range-queries with different selectivities that count the number of records. For each query selectivity, we executed queries with different range predicates to measure their actual I/O cost and report the average execution time. Figure \ref{fig:secondary-count_star} shows the execution times for both low and high selectivity predicates. For queries with low selectivity predicates, their execution times using the four different layouts were comparable, as shown in Figure \ref{fig:secondary_low}), and all queries took less than a second to finish. However, the execution times for queries that are 0.1\% selective were correlated with storage sizes (Figure \ref{fig:storage}). Figure \ref{fig:secondary_high} shows the execution times for queries with high selectivity predicates both with and without utilizing the $timestamp$ index. The secondary index accelerated the execution of queries with high selectivity predicates, except for the $AMAX$ layout. We observed that the scan-based query in the $AMAX$ layout (AMAX Scan) was faster to execute than its index-based queries. 

The previous experiment does not depict a full picture, as counting the number of records only accesses Page 0 of $AMAX$ and skips the rest of the pages. The benefit of using a secondary indexes for such queries become more apparent when a query accesses more columns. Figure \ref{fig:column-count_star} shows the impact of running queries that read different number of columns in the $APAX$ and $AMAX$ formats. Each query counts the appearances of different columns' values (i.e., non-NULL values) and varying the number of columns accessed from 1 to 10. The columns were picked at random and vary in terms of their types and sizes. Figure \ref{fig:colunm_time} shows the execution times for scan-based queries that access different number of columns. As expected, accessing more columns in the $AMAX$ format negatively impacts query performance, while the performance was relatively stable in the $APAX$. For example, reading ten different columns was 9.5X slower than reading a single column for the $AMAX$ layout, whereas the impact was less noticable in $APAX$. Despite the slowdown, querying records in the $AMAX$ layout was still faster than $APAX$. The time variance in accessing different columns shown in both figures is due to the time needed for reading and decoding values with different sizes. For example, change for reading six columns was higher than for seven columns, as the sixth column contained required values, while the seventh column's values were mostly NULLs. That was for the scan-based queries. Figures \ref{fig:column_1} -- \ref{fig:column_10} show the execution times of index-based queries with different selectivities (0.001\% -- 1.0\%). The execution times for all queries were comparable for both layouts, despite the number of columns each query reads. Compared to the scan-based queries, the index-based queries took less time to execute and were less sensitive w.r.t the number of columns. Thus, as for row-major layout, secondary indexes can accelerate queries against records in a columnar layout and can help to minimize the impact of reading multiple columns for $AMAX$-like layouts.
\section{Related Work}

\textbf{Columnar layouts with dynamic schema:}
Storing schemaless semi-structured data in a columnar layout has gained more traction lately, and several approaches have been proposed to address the issues imposed by schema changes. Delta Lake \cite{deltalake}, a storage layer for cloud object stores, addresses the challenges of updating and deleting records stored in Parquet files. Delta Lake recently added support for schema evolution; however, it still lacks support for storing heterogeneous values, as per Parquet's limitation. Alsubaiee et al. proposed a patented technique \cite{sattam_patent} that exploits Parquet's file organization to store datasets with heterogeneous values. The main idea of their approach is congregating records with the same value types within a group. In this work, we proposed an extension to Dremel to natively support union types, storing values with different types as different columns. Additionally, our extension to Dremel is independent of how the columns are organized --- making it a suitable format for different detailed storage organizations such as APAX, AMAX, or even Parquet's row groups. 

In \cite{json_tiles}, the authors have proposed \textit{Json Tiles}, a columnar format for semi-structured records integrated into Umbra \cite{umbra}, a disk-based column-store RDBMS. The proposed approach infers the structure of the ingested records and materializes the common parts of the records' values, including heterogeneous values, as \textit{JSON Tiles}. In our proposed work, we transform entire (nested and variant) records into columns. Additionally, we have targeted LSM-based systems, where in-place updates are not permitted, whereas \textit{JSON Tiles} are designed for systems where updates are performed in-place. 

For LSM-based document stores, Rockset \cite{rockset} supports storing values of semi-structured records in a columnar format, with the values of a column being stored in RocksDB \cite{rocksdb} (Rockset's storage engine) using a shared key prefix. To illustrate, let us consider an example of two records with the $id$s 1 and 2, where both records have additional fields $age$ and $name$. The keys $age.1$ and $age.2$ for the records with $id$s 1 and 2, respectively, share the same prefix $age$, which corresponds to the field $age$. Similarly, $name.1$ and $name.2$ share the same prefix $name$. When stored in RocksDB, the keys $age.1$ and $age.2$ and their values will be stored contiguously first, followed by the $name$'s values, as the RocksDB's ordering of keys dictates. Hence, when accessing either value, Rockset will only reads the required values from disk, which minimizes the I/O cost. This approach does not support encoding the column's values (e.g., via run-length encoding), however, which means more I/O.

\textbf{LSM-based column stores:} 
Most column-store databases employ a similar mechanism to LSM-based storage engines, where newly inserted records are batched in memory and then flushed to disk, during which time the flushed records are encoded and compressed. For example, Vertica \cite{c-store} and Microsoft SQL Server's column store \cite{sql_server_column1,sql_server_column2} employ an LSM-like mechanism, while column-store systems such Apache Kudu \cite{kudu} and ClickHouse \cite{clickhouse} are LSM-based. Update and delete operations may slightly differ from system to system, but they mostly share the objective of minimizing the cost of modifying columns' values in place. Deleted records are, therefore, usually marked and garbage-collected at a later stage, and updates are handled as a delete followed by an insert. This work is no exception, as we share similar challenges and objectives. Again, however, our focus is on nested and schemaless data.

\textbf{Code generation and query compilation:}
Data processing engines like Spark and DBMSs like Vector and Umber have moved from using the iterator model or the vectorized model for their query execution engines. Instead, they use code generation techniques to improve performance. Most such systems utilize strongly-typed languages for code generation, which is sufficient for schema-ful systems. However, for schemaless systems like MongoDB and Apache AsterixDB, utilizing a strongly-typed language would require adding additional checks to ensure the types of each processed value, which means more branches in the generated code. The Truffle framework addresses this issue for dynamically-typed languages such as Python and JavaScript. Earlier, Truffle was used to execute stored procedures and user-defined functions in the Oracle database and MySQL \cite{oracle_truffle_udf}. Recent work \cite{dynq} has proposed using the Truffle framework for code generation and query compilation to run Language-integrated Query (LINQ) over a dynamically-typed collection in JavaScript or R and showed that the performance of their approach was comparable to hand-written code. In this work, we have also used the Truffle framework, where we generate code for parts of a query plan in Apache AsterixDB, after which then the generated code is distributed and executed in parallel.


\section{Conclusion and Future work}
In this paper, we presented several techniques to store and query data in a columnar format for schemaless, LSM-based document stores. We first proposed several extensions to the Dremel format to make storing arrays' values more concise and to accommodate heterogeneous data values. Next, we introduced APAX and AMAX, two columnar layouts for organizing and storing records in LSM-based document stores. Furthermore, we highlighted the challenges involved reading and writing records in the APAX and AMAX layouts and proposed solutions to overcome those challenges. Experiments showed that the AMAX layout significantly reduced the overall storage overhead compared to the row-major formats. The impact of transforming records into columns during data ingestion varied according to the structure of thr ingested records, and it was seen that the AMAX layout's ingestion rate was relatively stable compared to APAX and faster as compared to AsterixDB's current schemaless format.

Additionally, we presented an approach for code generation and query compilation for schemaless document stores using the Truffle framework's JIT compilation capability to process values with heterogeneous types. Although we only generate code for a part of the query, our experiments showed a significant improvement over the current Hyracks execution model, even for row-major formats. In our evaluation, queries against the AMAX were the fastest to execute compared to other layouts, and for certain queries, the AMAX layout improved the query performance by orders of magnitude.

To the best of our knowledge, most column store databases, except Vector \cite{vector}, do not support secondary indexes, as scan-based queries are often considered good enough for data warehouse workloads. In this work, we evaluated the impact of secondary indexes on data ingestion and query performance for columnar formats and showed that the ingestion rate might be negatively impacted; however, the impact of reading multiple columns in AMAX was reduced when answering queries with secondary indexes.

In future work, we plan to extend the AMAX layout to support dictionary encoding, where we will dedicate different pages to store the values' dictionaries. Moreover, we plan to explore ways to merge dictionary pages efficiently during the LSM merge operation. Finally, we plan to expand our code generation framework to the entire quer to include pipeline-breakers such as group-by, order-by, and join operators.

%

\balance
\bibliographystyle{ACM-Reference-Format}
\bibliography{wail}
\clearpage
\appendix
\section{Queries}
\label{appen:queries}
In this section, we show the queries we ran in our experiments against the $cell$, $tweets\_1$, $sensors$, and $wos$ datasets.

\subsection{$cell$ Queries}
\label{appen:cell}

\begin{sqlpp}
Q1:
SELECT VALUE COUNT(*) 
FROM Tweets
\end{sqlpp}

\begin{sqlpp}
Q2:
SELECT caller, MAX(c.duration) as m
FROM Cell c
GROUP BY c.caller AS caller
ORDER BY a DESC
LIMIT 10
\end{sqlpp}

\begin{sqlpp}
Q3:
SELECT VALUE COUNT(*)
FROM Cell c
WHERE c.duration >= 600
\end{sqlpp}


\subsection{$tweets\_1$ Queries}
\label{appen:twitter}

\begin{sqlpp}
Q1:
SELECT VALUE COUNT(*) 
FROM Tweets
\end{sqlpp}

\begin{sqlpp}
Q2:    
SELECT VALUE uname,a
FROM Tweets t
GROUP BY t.users.name AS uname 
WITH a AS MAX(length(t.text))
ORDER BY a DESC
LIMIT 10
\end{sqlpp}

\begin{sqlpp}
Q3
SELECT uname, COUNT(*) as c
FROM Tweets t
WHERE (
  SOME ht IN t.entities.hashtags 
  SATISFIES LOWERCASE(ht.text) = "jobs"
)
GROUP BY user.name as uname
ORDER BY c DESC
LIMIT 10
\end{sqlpp}


\subsection{Sensors Dataset's Queries}
\label{appen:sensors}

\begin{sqlpp}
Q1:
SELECT VALUE COUNT(*)
FROM Sensors s, s.readings r
\end{sqlpp}

\begin{sqlpp}
Q2:
SELECT MAX(r.temp), MIN(r.temp)
FROM Sensors s, s.readings r
\end{sqlpp}

\begin{sqlpp}
Q3:
SELECT sid, max_temp
FROM Sensors s, s.readings as r
GROUP BY s.sensor_id as sid 
WITH max_temp as MAX(r.temp)
ORDER BY t DESC
LIMIT 10
\end{sqlpp}

\begin{sqlpp}
Q4:
SELECT sid, max_temp
FROM Sensors s, s.readings as r
WHERE s.report_time > 1556496000000 
AND s.report_time < 1556496000000 
                    + 24 * 60 * 60 * 1000
GROUP BY s.sensor_id as sid 
WITH max_temp as MAX(r.temp)
ORDER BY max_temp DESC
LIMIT 10
\end{sqlpp}

\subsection{$wos$ Queries}
\label{appen:wos}

\begin{sqlpp}
Q1:
SELECT VALUE COUNT(*)
FROM Publications as t
\end{sqlpp}

\begin{sqlpp}
Q2:
SELECT v, COUNT(*) as cnt
FROM Publications as t, 
   t.static_data.fullrecord_metadata
   .category_info.subjects.subject 
   	   AS subject
WHERE subject.ascatype = "extended"
GROUP BY subject.`value` as v
ORDER BY cnt DESC	
\end{sqlpp}

\begin{sqlpp}
Q3:
SELECT country, COUNT(*) as cnt
FROM (
  SELECT value countries
  FROM Publications as t
  LET  address = t.static_data
                 .fullrecord_metadata
                 .addresses.address_name,
       countries = ARRAY_DISTINCT(
              address[*].address_spec.country
              )
  WHERE IS_ARRAY(address)
  AND ARRAY_COUNT(countries) > 1
  AND ARRAY_CONTAINS(countries, "USA")
) as collaborators
UNNEST collaborators as country
WHERE country != "USA"
GROUP BY country
ORDER BY cnt DESC
LIMIT 10
\end{sqlpp}

\begin{sqlpp}
Q4:
SELECT pair, COUNT(*) as cnt
FROM (
    SELECT value ARRAY_PAIRS(countries)
    FROM Publications as t
    LET  address = t.static_data
                 .fullrecord_metadata
                 .addresses.address_name,
         countries = ARRAY_DISTINCT(
            address[*].address_spec.country
            )
    WHERE IS_ARRAY(address)
    AND ARRAY_COUNT(countries) > 1
) as country_pairs
UNNEST country_pairs as pair
GROUP BY pair
ORDER BY cnt DESC
LIMIT 10
\end{sqlpp}

\end{document}